\documentclass{aa}  

\usepackage{graphicx}
\usepackage[labelfont=bf]{subcaption}
\usepackage{txfonts}
\usepackage{mathtools}
\usepackage{hhline}

\usepackage{rotating}

\usepackage{hyperref} 
\hypersetup{colorlinks=true,citecolor=blue}
\usepackage{color}

\newcommand{\G}{\mathcal{G}}

\newcommand{\Mearth}{M_{\oplus}}

\newcommand{\Rp}{R_{\rm p}}
\newcommand{\dd}{\mathrm{d}}

\newcommand\Real{\operatorname{Re}}
\newcommand\Imag{\operatorname{Im}}

\usepackage[normalem]{ulem}
\usepackage{color}
\definecolor{blue}{RGB}{0,0,255}
\definecolor{red}{RGB}{255,0,0}
\definecolor{green}{RGB}{0,200,0}
\definecolor{black}{RGB}{0,0,0}

\begin{document}

   \title{Solid tidal friction in multi-layer planets: Application to Earth, Venus, a Super Earth and the TRAPPIST-1 planets}
   \subtitle{Can a multi-layer planet be approximated as a homogeneous planet?}

\titlerunning{Solid tides}

\author{E. Bolmont\inst{1,2} \and S. N. Breton\inst{2} \and G. Tobie\inst{3} \and C. Dumoulin\inst{3}  \and S. Mathis\inst{2} \and O. Grasset\inst{3}}

\offprints{E. Bolmont,\\ email: emeline.bolmont@unige.ch}

\institute{
$^1$ Observatoire de Gen\`eve, Universit\'e de Gen\`eve, 51 Chemin des Maillettes, CH-1290 Sauverny, Switzerland \\
$^2$ AIM, CEA, CNRS, Universit\'e Paris-Saclay, Universit\'e Paris Diderot, Sorbonne Paris Cit\'e, F-91191 Gif- sur-Yvette, France \\
$^3$ Laboratoire de Plan\'etologie et G\'eodynamique,UMR-CNRS 6112, Universit\'e de Nantes, 44322 Nantes cedex 03, France}

  \date{Submitted to A\&A}

 
  \abstract
   {With the discovery of TRAPPIST-1 and its seven planets within 0.06 au, the correct treatment of tidal interactions is becoming necessary.
   The eccentricity, rotation, and obliquity of the planets of TRAPPIST-1 are indeed the result of tidal evolution over the lifetime of the system.
   Tidal interactions can also lead to tidal heating in the interior of the planets (as for Io), which can then be responsible for volcanism and/or surface deformation.
   In the majority of studies to estimate the rotation of close-in planets or their tidal heating, the planets are considered as homogeneous bodies and their rheology is often taken to be a Maxwell rheology.  
  
   We investigate here the impact of taking into account a multi-layer structure and an Andrade rheology on the way planets dissipate tidal energy as a function of the excitation frequency.
   We use an internal structure model, which provides the radial profile of structural and rheological quantities (such as density, shear modulus and viscosity) to compute the tidal response of multi-layer bodies. 
   We then compare the outcome to the dissipation of a homogeneous planet (which only take a uniform value for shear modulus and viscosity). 
   
   We find that for purely rocky bodies, it is possible to approximate the response of a multi-layer planet by that of a homogeneous planet. 
   However, using average profiles of shear modulus and viscosity to compute the homogeneous planet response leads to a huge overestimation of the averaged dissipation. 
   We provide fitted values of shear modulus and viscosity to be able to reproduce the response of various types of rocky planets.
   However, we find that if the planet has an icy layer, its tidal response can no longer be approximated by a homogeneous body because of the very different properties of the icy layers (in particular their viscosity), which lead to a second dissipation peak at higher frequencies. 
   We also compute the tidal heating profiles for the outer TRAPPIST-1 planets (e to h). 
   }


   \maketitle
%

\section{Introduction}

As we are discovering Habitable Zone (HZ) planets around low-mass stars (e.g. TRAPPIST-1: \citealt{2016Natur.533..221G,2017Natur.542..456G} and Proxima-b: \citealt{2016Natur.536..437A}), the correct modeling of tides is becoming increasingly mandatory in orbital dynamics studies and in order to understand the feedback between thermal and tidal evolution of these exoplanets.
Indeed, these planets are sufficiently close-in and tidal interactions can play a major role \citep[e.g.][]{2018haex.bookE..24M}. 
In particular, tidal effects impact the evolution of the rotation and obliquity of the planets and on a longer timescale their eccentricity and semi-major axis.
They can also significantly contribute to their internal heat budget through tidal friction processes (as for Io: \citealt{2000Sci...288.1198S}; or for HZ planets around brown dwarfs: \citealt{2018haex.bookE..62B}).
Rotation and obliquity have a key influence on the climate of planets (heat redistribution and seasonal effects, respectively).
These quantities are known to have influenced Earth's climate \citep[the Milankovitch cycle, which can be reproduced by numerical integrations, see][]{1993A&A...270..522L}.
As these quantities are not currently retrievable from observations, we need a correct tidal theory to try to estimate them and be able to say if a given planet is more likely to be tidally locked (solution favored for low eccentricities) or in spin-orbit resonance (for higher eccentricities, e.g. \citealt{2013ApJ...764...27M}).
To be able to do this would require a precise knowledge of the system's parameters, such as mass, radius, eccentricity.
In some circumstances, tides may induce significant melting in the interior and enhance volcanic activity, possibly favoring lithospheric weakening and plate generation \citep{2019Icar..325...55Z}.
While the theoretical end of the tidal evolution for a single planet is known (if the orbital angular momentum is more than 3/4 of the total angular momentum of the system, the following equilibrium is reached: the orbit of the planet should be circular, its rotation synchronized and its spin aligned with the orbital angular momentum vector, see \citealt{1980A&A....92..167H}), the evolution towards that tidal equilibrium, especially the timescale at which the final state is reached, depends on the way the planets dissipate tidal energy.
The situation is different in a multi-planet system, where the value of the eccentricity will be the result of the competition between tidal damping and planet-planet interactions \citep[e.g.][]{2013A&A...556A..17B}.
Depending on this eccentricity, the rotation could be different than synchronization. 

\begin{figure}     
\begin{center}
\includegraphics[width=0.7\linewidth]{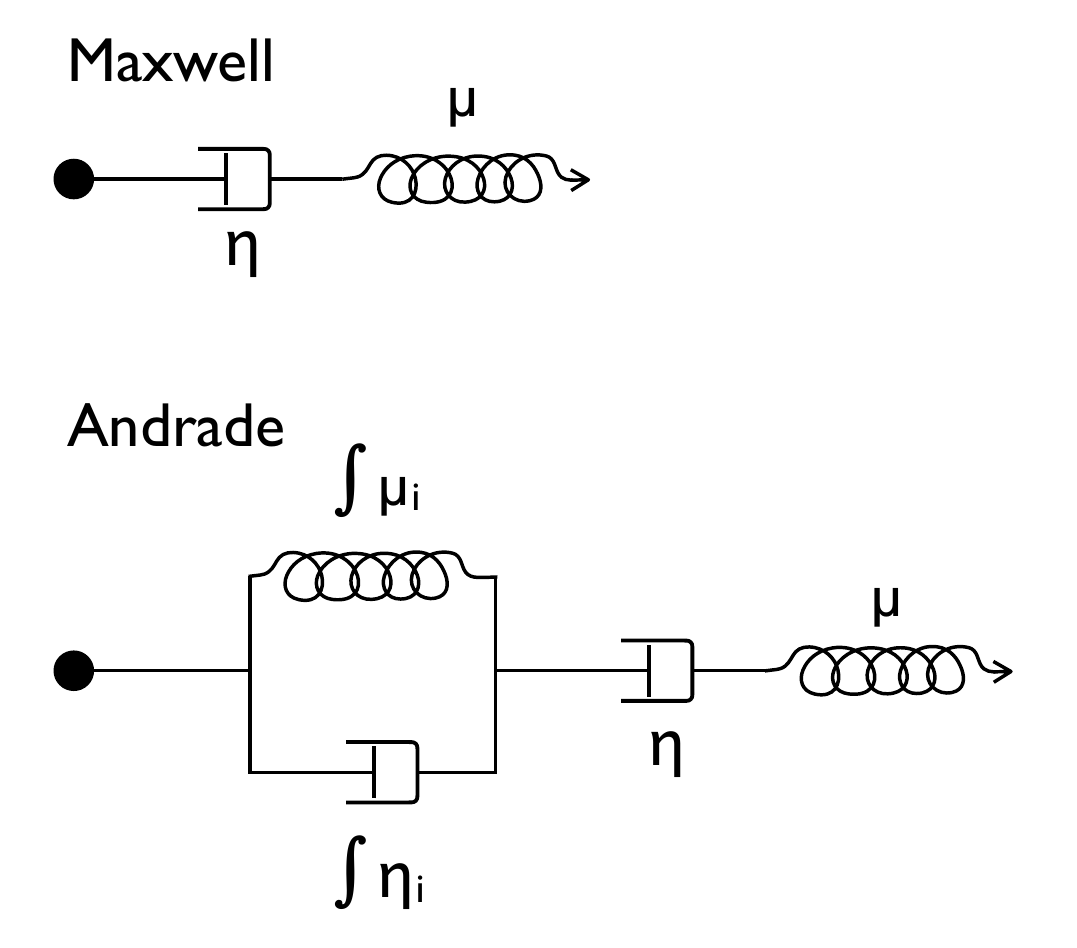}
\caption{Schematic representation of the Maxwell viscoelastic model and the Andrade anelastic model. The different models can be seen as a succession of springs (labeled as $\mu$) and dashpots (labeled as $\eta$). The springs represent the elastic properties of the material while the dashpots represent its viscous properties. The Andrade model is a combination of the viscoelastic Maxwell assembly in series with an infinite number of springs and dashpots in parallel. This last part represents the memory of the material, which corresponds to the anelastic component of the assembly.}
\label{Maxwell_vs_Andrade}
\end{center}
\end{figure}

Traditionally for tidal orbital evolution studies, simple equilibrium tide models are used like the Constant Time Lag model (CTL model, \citealt{1979M&P....20..301M}, \citealt{1981A&A....99..126H}, \citealt{1998ApJ...499..853E}) or Constant Phase Lag model (CPL model, \citealt{1966Icar....5..375G}).
The CTL model consists in assuming that the deformable body is made of a weakly viscous fluid \citep{1973Ap&SS..23..459A}.
In this framework, while the eccentricity is non zero, the planetary rotation tends to the pseudo-synchronous rotation \citep{1981A&A....99..126H}.
However, following the formalism of \citet{2012CeMDA.112..283E} and \citet{2012ApJ...746..150E}, \citet{2013ApJ...764...27M} showed that for rocky planets, the hypothesis of weakly viscous fluid is no longer valid and a rheology more appropriate for a rocky planet should be considered.
In this particular example, a combined rheological model was used (\citealt{1910RSPSA..84....1A} at higher frequencies and Maxwell at lower frequencies) to show that the planets are trapped in spin-orbit resonances, whose order depends on the eccentricity (like Mercury, \citealt{2014Icar..241...26N}).
The rheology of Maxwell and Andrade are widely used, for instance to estimate rotation states \citep[e.g.][]{2014A&A...571A..50C} or tidal heating in planets \citep[e.g.][]{2013ApJ...764...27M,2018ApJ...857...98R,2019A&A...630A..70T}.
Maxwell's rheology describes a viscoelastic material while Andrade's rheology describes an anelastic material.
Figure~\ref{Maxwell_vs_Andrade} shows the schematics of each models; both can be modeled as a succession of springs and dashpots \citep[e.g.][]{2011JGRE..116.9008C}.
While the Maxwell model is simpler, Andrade's model is known to better reproduce the response of materials at higher frequencies \citep[][]{2007JGRE..11212003E, 2011JGRE..116.9008C}. 
The response of various materials has been experimentally tested over the years for rocks \citep{2003PCM....30..157W,2004JGRB..109.6201J,2010PMag...90.2817S} and ices \citep{2013ASSL..356..183M,2015GeoRL..42.6261C}. 
These studies have shown that the Andrade model performs fairly well, although it is sometimes necessary to use more complex rheologies still \citep[e.g.][]{2010PMag...90.2817S}.

Taking into account a better description of the tidal deformation was done to estimate the rotation states of planets \citep[e.g.][]{2013ApJ...764...27M,2014Icar..241...26N}, their rotational and orbital evolution \citep[e.g.][]{2014A&A...571A..50C,2016CeMDA.126...31B,2016MNRAS.458.2890F} or to estimate the tidal heating in rocky planets (e.g. \citealt{2009ApJ...707.1000H,2014ApJ...789...30H,2019A&A...630A..70T} for generic planets and \citealt{2018A&A...613A..37B} for TRAPPIST-1).  
Most of these different studies are considering a homogeneous body \citep[except for][]{2014ApJ...789...30H,2019A&A...630A..70T} and most of them are assuming a Maxwell rheology \citep[except for][]{2013ApJ...764...27M,2014Icar..241...26N,2019A&A...630A..70T}.
Note that \citet{2017CeMDA.129..235W} revisited the work of \citet{2014A&A...571A..50C} using a multi-layer model for the planets and investigated the tidal dissipation and tidal torques for the Maxwell and the Andrade rheology.
They showed that considering a multi-layer structure does not significantly impact the rotational states of the planets.

%
%
%

In this article, we investigate the frequency dependency of tidal dissipation for multi-layer bodies and the resulting tidal heating profiles in rocks and ices and compare our results with those obtained with the formulation proposed for a homogeneous body by \citet{2012ApJ...746..150E}.
We do not consider here the contribution of potential liquid layers, indeed the tidal response of liquid layers is much more complex to account for, as the equilibrium tide is accompanied by the dynamical tide.
The dynamical tide would consist here of gravito-inertial waves which are excited by the perturber and the resulting frequency dependence of the dissipation is more erratic \citep[see][]{2004ApJ...610..477O,2015A&A...581A.118A,2018A&A...615A..23A}.
As a first step, we therefore neglect these layers, which could be a sub-surface or surface ocean \citep[as could be present on the surface of TRAPPIST-1e, see][]{2018A&A...612A..86T}, or melted regions in the interior (for example, due to radiogenic or tidal heating, see \citealt{2009ApJ...707.1000H} for the latter).
In Section~\ref{models_dissipation}, we present the methods we use to calculate the tidal dissipation for both: homogeneous planets and multi-layer planets. 
In order to calculate the dissipation of a multi-layer planet, we need an internal structure model, which give us the profiles for relevant quantities (such as shear modulus and viscosity). 
We use here two internal structure models, which we present in Section~\ref{structure_model}.
We consider different types of planets: an ocean-less Earth-like planet, a Venus-like planet, three ocean-less rocky planets of $0.5$, $5$ and $10~\Mearth$, and the outer planets of the TRAPPIST-1 system (from e to h).
We concentrate here on the outer planets of TRAPPIST-1 to be able to safely neglect potential melted regions in the planets.
To determine the internal structures of the outer TRAPPIST-1 planets, we used the most probable mass determined by \citet{2018A&A...613A..68G}. 
We then made assumptions on the iron over silicate ratio and inferred the radius which is within the observational uncertainty range of \citet{2018MNRAS.475.3577D}. 
This article does not aim at precisely characterizing the internal structure of the TRAPPIST-1 planets, but aims at drawing general conclusions on the dissipation in the interior of multi-layer planets. 
We therefore did not explore the whole range of allowed masses and radii.
When the masses of the TRAPPIST-1 planets will be sufficiently refined, a dedicated study on this system will be justified.
For Venus, we choose here to use a more specific model to evaluate the tidal dissipation.
Knowing the dissipation of the rocky part of Venus is a necessary step to study the equilibrium rotation of the planet, as dissipation of solid body tides controlled the first stage of planetary despinning, before atmospheric friction and thermal atmospheric tides played a dominant role \citep[e.g.][]{2001Natur.411..767C,2015Sci...347..632L,2017A&A...603A.108A}.
Indeed the rotation of Venus is determined by  the balance between the gravitational tide which acts to synchronize the rotation and the atmospheric tide which acts to de-synchronize the rotation \citep{1978Natur.275...37I,1980Icar...41....1D,2017A&A...603A.107A}.
Estimating as precisely as possible the gravitational tide is therefore important to understand Venus' rotation.
In Section~\ref{dissipation_vs_frequency}, we show how the dissipation of multi-layer planets vary with the excitation frequency and compare it with the dissipation of homogeneous planets.
For the different types of planets, we provide fitted values of shear modulus and shear viscosity which allow to best reproduce the dissipation of a multi-layer planet with that of a homogeneous planet.
Finally in Section~\ref{tidal_heating_T1}, we compute the tidal heating profiles of the outer (multi-layer) planets of TRAPPIST-1.



\section{Models of dissipation}\label{models_dissipation}

Let us consider two bodies, one is a point mass, which we call the perturber P and the second one is an extended central mass C, which is going to respond to the perturber's gravitational potential.
Let us call $M_P$ and $M_C$ the respective masses of P and C. 
The radius of C will be referred to as $R_C$.
The excitation frequency will be referred to as $\omega$, or $\omega_{lmpq}$ where $l, m, p, q$ are the indices of the harmonic expansion of the potential\footnote{The generic form of the excitation is: $\omega_{lmpq} = (l-2p+q)n-m\Omega$, where $n$ is the mean motion of the planet and $\Omega$ is its spin. For a circular coplanar orbit, $\omega = \omega_{2200} = 2(n-\Omega)$.}.
We consider that the perturber is farther away than $5\times R_C$ from the central mass and that the eccentricity is low enough, so we can restrict the expansion of the tidal potential and perturbations on spherical harmonics to the $l=2$ mode \citep[e.g.][]{2009A&A...497..889M,2013ApJ...764...27M}.

The Love number \citep{1911spge.book.....L} quantifies the ability of a celestial body to respond to tidal forcing. 
It corresponds to the ratio between the additional gravitational potential induced by internal mass redistribution due to tidal deformation and the external tidal potential created by the perturber  
\begin{equation}
k_2(\omega) = \frac{\Phi_{\rm deformed~body}(r = R_C)}{\Phi_{\rm perturber}(r = R_C)}.
\end{equation}
For a perfectly elastic body, $k_2$ is a real quantity, the deformation is instantaneous and the tidal bulges are aligned with the direction of the perturber.
There is no tidal evolution in this case. 
However, for a real body, the response is never perfectly elastic and part of the response is dissipative, resulting in a delay/lag in the deformation.
$k_2$ becomes a complex quantity and its imaginary part quantifies the corresponding lag $\delta_2$ \citep{2013ApJ...764...26E}
\begin{equation}\label{love_num_delta}
\Imag k_2(\omega) = - \sin\delta_2(\omega) |k_2(\omega)|.
\end{equation}

The aim is here to quantify the amplitude of $k_2$ and the lag consistently with the internal structure and rheology of various rocky planets. 
We first discuss the case of a homogeneous planet and then move on to multi-layer planets.

\subsection{Dissipation of a homogeneous body}
\label{Homogeneous_body}

\begin{figure*}     
\centering
\includegraphics[width=0.95\linewidth]{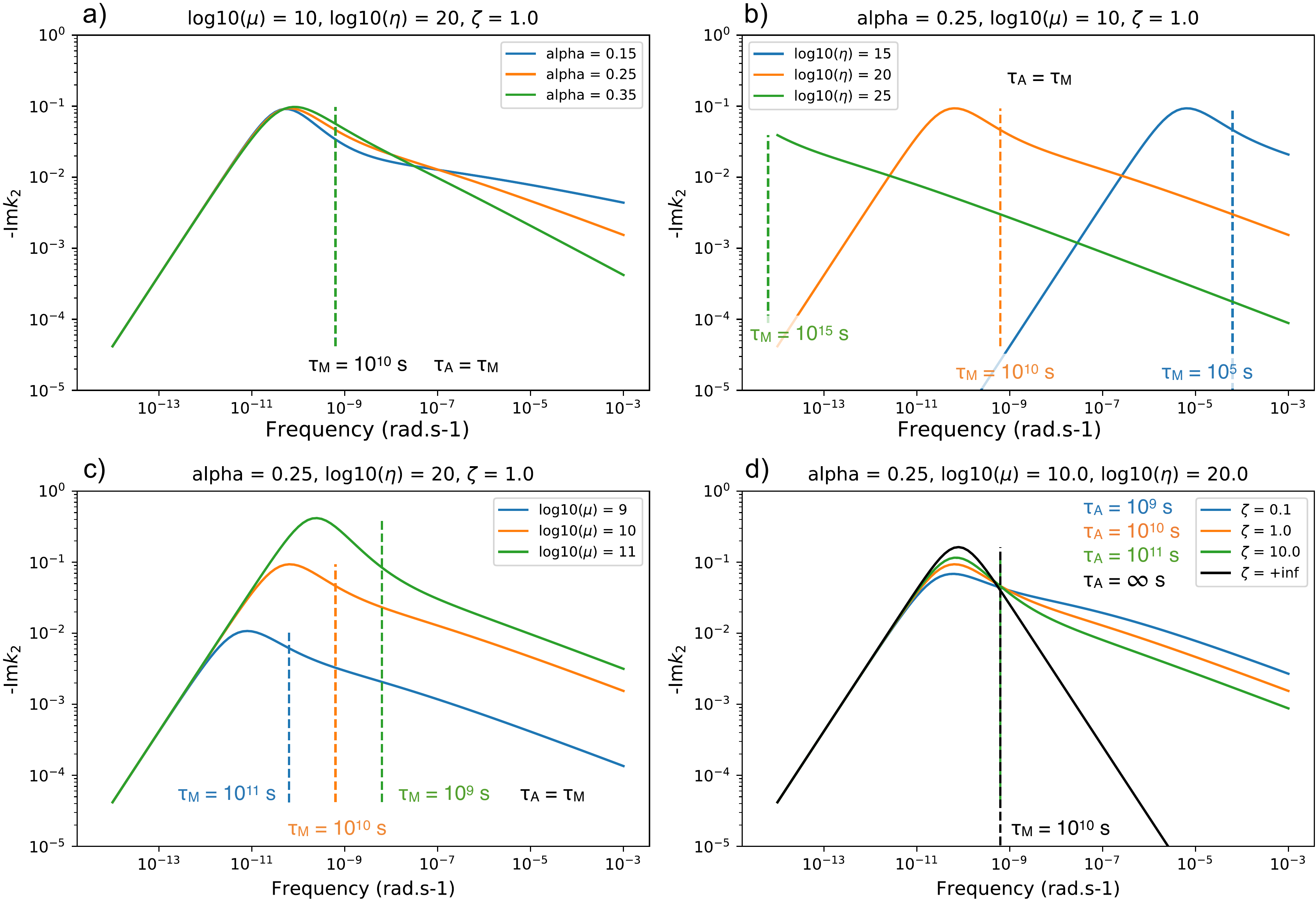}
\caption{Dissipation as a function of the excitation frequency for a homogeneous body. a) Effect of the parameter $\alpha$. b) Effect of the viscosity $\eta$. c) Effect of the shear modulus $\mu$. d) Effect of the parameter $\zeta$ (Note that a value $\zeta = +\infty$ corresponds to a Maxwell model).}
\label{Imk2_homog_parameters}
\end{figure*} 

Following \citet{2012ApJ...746..150E} we can express the imaginary part of the Love number as follows
\begin{flalign}
\Imag[k_{2, {\rm hom}}] = & \frac{3}{2} \frac{A_2 J \Imag[\bar{J}(|\omega|)] \times {\rm sgn}(\omega)}{\left(\Real[\bar{J}(|\omega|)]+A_2 J\right)^2 + \left(\Imag[\bar{J}(|\omega|)]\right)^2},\\
A_2 = & \frac{57}{8} \frac{J^{-1}}{\pi\G\rho_{\rm p}^2\Rp^2}, 
\end{flalign}
where $\bar{J}$ is the complex compliance of the material (in 1/Pa) and $\omega$ is the excitation frequency. 
For the Maxwell rheology model, the simplest viscoelastic model consisting of a dashpot (viscous element) and a string (elastic element) in series (Fig.~\ref{Maxwell_vs_Andrade}), the complex compliance is given by : 
\begin{equation}\label{J_Max}
\bar{J}(\omega)=\frac{1}{\mu_E}-\frac{i}{\eta\omega},
\end{equation}
where $\mu_E = 1/J$ is the (unrelaxed) elastic shear modulus, in Pa, (represented by the spring in Fig. \ref{Maxwell_vs_Andrade}) and $\eta$ is the shear viscosity, in Pa.s, (represented by the dashpot in Fig.\ref{Maxwell_vs_Andrade}). 
For a rheology of Andrade, the complex compliance is given by \citep{2011JGRE..116.9008C}
\begin{equation}
\bar{J}(\omega) = J - \frac{i}{\eta\omega} + \beta(i\omega)^{-\alpha}\Gamma(1+\alpha),
\label{J_And}
\end{equation}
where the third term describes the transient anelastic response, which controls the behavior at forcing periods comparable to the Maxwell time, defined as $\tau_M = \eta/\mu_{\rm E} = \eta J$.
For the sake of simplicity, we later refer to $\mu_{\rm E}$ as $\mu$.
$\alpha$ is a parameter linked to the duration of the transient response in the primary creep.  
This parameter is commonly considered to be between 0.20 and 0.40 \citep{2011JGRE..116.9008C} based on existing laboratory constraints on olivine minerals and ices.
Besides, \citet{2019A&A...630A..70T} showed that a parameter value between 0.23 and 0.28 allowed to well reproduce the dissipation factor of the present-day solid Earth. 
$\beta$ describes the intensity of anelastic friction in the material and is given by
\begin{equation}\label{beta}
\beta = J \tau_A^{-\alpha},
\end{equation}
where $\tau_A$ is the timescale associated to the Andrade creep, later referred as the Andrade time. 
The quantity $\beta$ can be expressed with the dimensionless parameter $\zeta$ defined as $\zeta = \tau_A/\tau_M$
\begin{equation}\label{beta1}
\beta = \zeta^{-\alpha} J \tau_M^{-\alpha}.
\end{equation}
\citet{2011JGRE..116.9008C} showed that under low stress $\zeta \approx 1$.
To compute the dissipation we fix $\zeta=1$ but to fit the parameters of a homogeneous planet to the response of a multi-layer planet (see Section~\ref{dissipation_vs_frequency}), we treat it as a free parameter.

At tidal periods much shorter than $\tau_M$ (high frequency), the response is dominated by the elastic behaviour, while at tidal periods much longer, it is dominated by viscous behaviour.  
Compared to the classically used Maxwell rheology, the Andrade rheology provides a more accurate description of the frequency dependency of both elastic and dissipative responses on a wide range of forcing frequencies/periods.  
The Maxwell rheological model provides a reasonable description of viscoelastic behavior for forcing periods near and larger than to the Maxwell time, but strongly underestimates viscous dissipation (or overestimates Q factor) for forcing periods much smaller than the Maxwell time \citep[e.g.][]{2012ApJ...746..150E,2019A&A...630A..70T}.\\

To sum up, to calculate the tidal response of a homogeneous body of given mass and radius, one needs effective values of the shear modulus $\mu$, the shear viscosity $\eta$, $\alpha$, $\zeta$ parameters representative of the whole planet interior.  
Figure~\ref{Imk2_homog_parameters} shows the dependence of the imaginary part of the Love number with the excitation frequency for different values of these four parameters.
Figure~\ref{Imk2_homog_parameters}a) shows that parameter $\alpha$ influences the behavior at high frequencies: the higher $\alpha$, the steeper the slope.
Figure~\ref{Imk2_homog_parameters}b) shows that changing the viscosity at constant shear modulus leads to a shift of the frequency at which the maximum dissipation occurs.
The maximum dissipation occurs at a frequency which is $1/\tau_M = \mu/\eta$ so that increasing the viscosity leads to a frequency shift of the dissipation maximum towards the small frequencies.
Figure~\ref{Imk2_homog_parameters}c) shows that the shear modulus impacts both the intensity of the maximum of dissipation and its frequency position. 
The higher the shear modulus (or the smaller the Maxwell Time), the higher the maximum dissipation and the higher the frequency corresponding to the maximum dissipation.
Finally changing the ratio of Andrade Time over Maxwell Time modifies the value of the maximum dissipation (see Fig.~\ref{Imk2_homog_parameters}d) and the slope occurring around a frequency of $10^{-9}$~rad.s$^{-1}$.
The higher $\zeta$, the more pronounced the peak.

\subsection{Dissipation in a multi-layer body}\label{multilayer_body}

To calculate the distribution of tidal dissipation in a spherical multilayer body, we use the elastic formulation of spheroidal oscillation developed by \citet{1972MetComPhy...1..217S}, adapted to the viscoelastic case by \citet{2005Icar..177..534T}, using the correspondence principle \citep{1954JAP....25.1385B}, and recently adapted to multilayer solid exoplanets \citep{2019A&A...630A..70T}. \\


Modeling the tidal response of a multi-layered interior consists of determining the displacements, stresses and potential perturbations induced by an external potential at any point in the body by solving the equation of motions and Poisson equations with appropriate boundary conditions and by considering a given stress-strain relationship, representative of the materials composing the  planetary interior. 
By assuming a spherically symmetric layered interior (i.e. no lateral variations in material properties is allowed), the problem solutions can be separated into a radial component and an orthoradial component using the spherical harmonics basis.\\

Following the approach of \citet{1972MetComPhy...1..217S}, the spheroidal deformations of a spherically symmetric layered interior can be formulated by using six radial functions, $y_i$, that satisfy a set of differential equations
\begin{equation} \label {Takeushi}
\frac {\mathrm{d}y_i(r, \omega_{lmpq})}{\mathrm{d}r} = \sum_{i = 1}^{6}\mathrm{\textbf{A}}_{ij} y_j (r, \omega_{lmpq}).
\end {equation}  
The matrix $\mathrm{\textbf{A}}_{ij}$ depends on the excitation frequency $\omega_{lmpq}$, on the density profile $\rho(r)$, on the compressibility modulus $\kappa_{\rm E}(r)$ and on the shear modulus $\mu_{\rm E}(r) = \mu(r)$ in the elastic case, and on the acceleration due to gravity. 
The functions $y_1$ and $y_3$ are respectively associated to the radial and tangential displacement, $y_2$ and $y_4$ to the radial and tangential stresses. 
The fifth function $y_5$ is associated to the gravitational potential and $y_6$ allows to insure the continuity of the gradient of the gravitational potential and is defined by
\begin {equation} \label{Eq21}
\begin{split}
y_6(r, \omega_{lmpq}) &= \frac {\mathrm{d}y_5(r, \omega_{lmpq})}{\mathrm{d}r} - 4\pi G \rho y_1(r, \omega_{lmpq}) \\
& \quad+ \frac {l+1}{r} y_5(r, \omega_{lmpq}).
\end{split}
\end {equation}
At the planet surface, the potential Love number of degree $l$ is given by
\begin {equation} \label{Lovenumber}
k_l (\omega_{lmpq}) = y_5(R_p, \omega_{lmpq}) - 1
\end {equation}
with $\Rp$ the radius of the body.
We are here considering only the degree-2 tidal potential \citep[it is enough if the perturber is more than $5~\Rp$ from the considered body and the eccentricity is low, e.g.][]{2009A&A...497..889M, 2013ApJ...764...27M}.
In the following, for simplicity we will refer to $\omega_{lmpq}$ as $\omega$.
For a purely elastic body, the relationship between the stress tensor and the strain tensor is determined by a Hooke's law.  \citep[e.g.][]{1973pedp.book.....M}.


We assume here a viscoelastic rheology and use the correspondence principle from \citet{1954JAP....25.1385B}. 
This principle states that for the same initial condition and geometry, the formulation of the elastic problem is equivalent to the formulation of the viscoelastic problem when the rheological parameters and the previously defined radial functions are complex. 
This means that the relation between the strain $\epsilon_{i,j}$ and stress $\sigma_{i,j}$ tensors for elastic bodies can be generalized for viscoelastic bodies, as a generalized Hooke's law in the frequency domain
\begin {equation} \label{Hooke_general}
\tilde\sigma_{i,j} = 2\tilde\mu(\omega)\tilde\epsilon_{i,j} + \left(\tilde\kappa - \frac{2}{3}\tilde\mu(\omega)\right)\tilde\epsilon_{k,k}\delta_{i,j},
\end {equation}
where the tilde indicates a complex value in the frequency domain, $\delta_{i,j}$ is the Kronecker's function, $\tilde\mu$ is a complex shear modulus and $\tilde\kappa$ is a complex compressibility modulus.
For a Maxwell rheological model, the complex shear modulus, which corresponds to the inverse of the complex compliance given in Eq. \ref{J_Max}, is
\begin {equation} \label{Hooke_general_mu}
\tilde\mu(\omega) = \frac{\mu(r)\omega^2\eta^2}{\mu^2 + \omega^2\eta^2} + i\frac{\mu^2\omega\eta}{\mu^2 + \omega^2\eta^2},
\end {equation}
which depends on the excitation frequency $\omega$ but also the radial coordinate $r$, via the radial dependence of the elastic shear modulus $\mu(r)$ and the viscosity $\eta(r)$. 
For the Andrade rheology, in a similar manner, the complex shear modulus is just the inverse of complex compliance given in Eq. \ref{J_And}.\\

The correspondence principle is equivalent to considering the equations governing the system but replacing the $y_i$ functions by their complex counterparts $\tilde y_i$, the elastic shear modulus $\mu$ by the complex $\tilde \mu$, the compressibility modulus $\kappa_E$ by $\tilde\kappa$ and the Love number of Eq.\ref{Lovenumber} by $\tilde k_2$.
The imaginary part of these quantities governs the dissipative response of the planet while the real part describe the instantaneous elastic response.
The imaginary part of the complex shear modulus is controlled by the shear viscosity $\eta(r)$, whereas the complex bulk modulus depends on the bulk viscosity, $\eta_b(r)$. As the bulk viscosity is poorly constrained for planetary materials \citep[e.g.][]{2001JGR...106.8887B}, the dissipative part of $\tilde{\kappa}$ is often neglected and only the elastic, real part is considered \citep[e.g.][]{2005Icar..177..534T}.

\medskip

In order to calculate the dissipation of a multi-layer body we therefore need the profiles of the density, bulk modulus, shear modulus and shear viscosity, as well as a rheological model to compute the complex shear modulus from the above cited rheological parameters assumed in each layer. 
The complex compliance is computed using Eq.~\ref{J_Max} or \ref{J_And} depending if a Maxwell or Andrade rheology is assumed. 
For the sake of simplicity, we assume the same rheological models in all internal layers. 
In the case of the Andrade model, we also consider the same constant parameters $\alpha$ and $\zeta = \tau_A/\tau_M = 1$ for all internal layers.

We calculate the response of different bodies within this formalism for three different $\alpha$ parameters: 0.15, 0.25 and 0.35, which bracket the reasonable parameter space \citep{2011JGRE..116.9008C}. 
Finally, for the profiles of the different quantities (density, compressibility modulus, shear modulus and viscosity) we use the following internal structure model.

\subsection{Internal structure model}\label{structure_model}

For the planets we consider in this study (a planet of $0.5~\Mearth$, an Earth-like planet, 2 Super-Earths of $5~\Mearth$ and $10~\Mearth$, and TRAPPIST-1e), we obtain the radial profiles of the different quantities following the formalism of \citet{2007Icar..191..337S}. 
We present here the main ingredients and characteristics of the internal structure model and we refer the reader to \citet{2007Icar..191..337S} for the details.
We also study the case of a Venus-like planet, for which we use a more specific model based on \citet{2017JGRE..122.1338D}.

\subsubsection{Generic model from \citet{2007Icar..191..337S}}\label{structure}

\begin{figure}[bhtp]     
\begin{center}
\includegraphics[width=\linewidth]{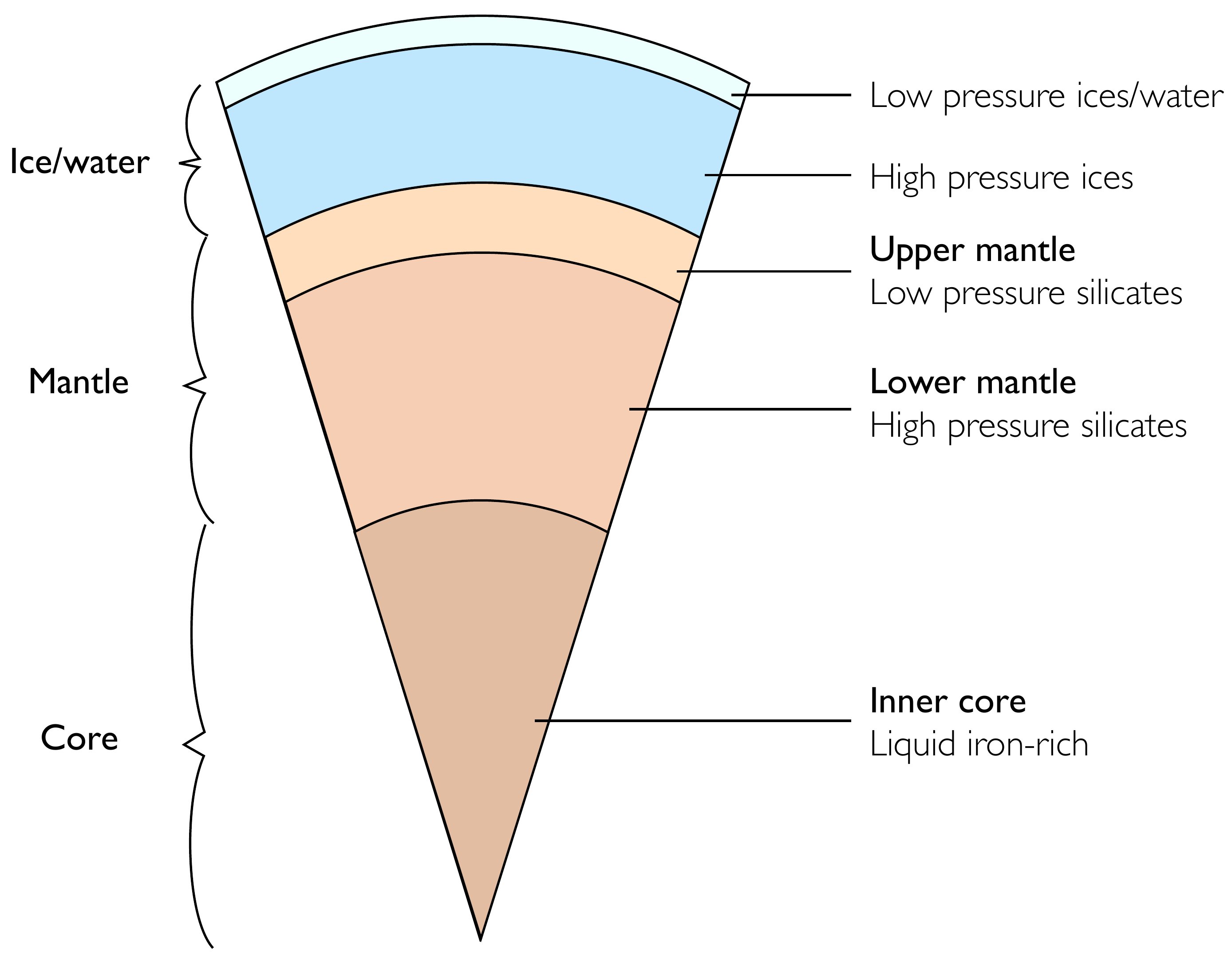}
\caption{Internal structure model.}
\label{internal_structure_model}
\end{center}
\end{figure}

The most generic model is composed of 5 layers: one layer representing the inner core, two layers for the mantle, and two layers for ices and liquid water on top.
Figure~\ref{internal_structure_model} shows the set up. 

In this model, the first and innermost layer is a liquid iron-rich core. 
For the sake of simplicity, no solid inner iron core is considered, as it has only little effect on the global tidal deformation \citep[][]{2017JGRE..122.1338D,2019A&A...630A..70T}.

The mantle is divided into two layers because of a mineralogical transformation that occurs at a pressure of about 25~GPa. 
Similar to the Earth's lower mantle, the lowermost part of the silicate mantle is composed of high-pressure silicate minerals: 
perovskite and magnesiow\"ustite\footnote{Note that we do not take into account here the post-perovskite phase (high-pressure phase of perovskite). While taking into account this phase should not change the mass-radius relationship of the planets, it could have an impact on the tidal heating through the viscosity and shear modulus.}. 
The upper mantle (third layer)  is composed of low pressure silicate minerals: olivine, ortho- and clino-pyroxenes and garnet.
For ice-rich exoplanets, two additional layer are considered: high-pressure ice layer when the pressure exceeds 2.2 GPa (fourth layer) and low-pressure ice layer or water layer  (fifth layer) depending on the surface temperature.
This outer layer is very thin and does not contribute significantly to the mass-radius relationship of the planet \citep{2007Icar..191..337S}. 
However, its physical state (liquid or solid) can significantly affect the tidal deformation and dissipation \citep[e.g.][]{2018A&A...615A..23A,2019A&A...629A.132A} and therefore its thickness and state need to be carefully determined to correctly predict the tidal response.

Following the approach of \citealt{2007Icar..191..337S}, a Birch-Murnagan Equation of State (EoS), up to the third order in finite strain, is considered for the upper mantle and the low-pressure ice layer or water layer (layers 3 and 5), whereas a Mie-Grun\"eisen-Debye is employed for the iron core, the lower silicate mantle and the high-pressure ice mantle (layers 1, 2, 4).\\

The elastic bulk isentropic modulus, $K$, is derived from the density and pressure profiles 
\begin{equation}
K=\rho\frac{dP}{d\rho}.
\end{equation}
The shear modulus, $\mu$, in each solid layer is then estimated from the bulk modulus and the pressure. 
For the silicate part, the following relationship, which reproduces well the shear modulus profile in the Earth's mantle \citep{1981PEPI...25..297D,2008phea.book.....S}, is used
\begin{equation}
\left(\frac{\mu}{K}\right)_{sil}=0.52-0.5\frac{P}{K} 
\end{equation}
for $P< 25$ GPa, and 
\begin{equation}
\left(\frac{\mu}{K}\right)_{sil}=0.63-0.885\frac{P}{K}
\end{equation}
for $P> 25$ GPa. 
For pressure above 130~GPa, we neglect the phase transition to the post-perovskite and thus consider the same relationship. 
For the ice layers, a similar relationship constrained from existing experimental data at high pressures \citep{1983PhRvB..27.6409P} is used
\begin{equation}
\left(\frac{\mu}{K}\right)_{ice}=0.6-0.9\frac{P}{K}.
\end{equation}
Note that $\mu/K=0.6$ corresponds to a Poisson body.\\

The internal structure is computed from the assumed planet composition which is defined using four parameters: the water mass relative to the mass of the planet, the bulk ratios of  Mg/Si and Fe/Si, and the Mg content of the silicate mantle, defined as the Mg number, Mg\# (the mole fraction Mg/(Mg+Fe) in the silicates). 
We also need to give three additional constraints: the composition of the core is fixed (here to $87\%$ of Fe and $13\%$ of FeS), all the water is found in layers 4 and 5 (no water incorporated in the silicates) and the mantle is chemically homogeneous.
Knowing the former four parameters combined with the latter three assumptions, both the size of each layer and the mineralogical composition of the silicates mantles can be determined.


For the Earth, for instance, there is no layer 4 (no high pressure ices) and layer 5 is so thin that it does not contribute to the mass-radius relationship. 
The Earth-like planet we consider here is an ocean-less planet\footnote{Note that Earth's ocean are responsible for most of the dissipation \citep[via the complex interaction of the oceans with the rocky surface, see][]{1977RSPTA.287..545L, 2000Natur.405..775E}} of Earth's radius and mass so that layer 5 does not exist either.  
In this model, Earth is therefore made up of three layers, the upper mantle, the lower mantle and the liquid core (note that we neglect here the Earth's small solid core as it represents only 2\% of Earth's mass). 
This model reproduces the Preliminary Reference Earth Model (PREM, \citealt{1981PEPI...25..297D}) as shown in \citet{2007Icar..191..337S}.
For the Earth-like planet and the $0.5~\Mearth$ planet and the 2 super-Earths, we also vary the Fe/Si ratio compared to the Earth's ratio.  
We consider three different Fe/Si contents compared to the Earth’s: $50\%$, $100\%$ and $150\%$.
For a constant mass, increasing the quantity of iron leads to a denser and smaller planet.
Table~\ref{table1} summarizes the assumed parameters for the different planets considered in this study.

Viscous dissipation is generated in the solid layers in which the viscosity and shear modulus are non zero.
The liquid core and the water ocean (when present) are assumed to be inviscid, which means that there is no viscous dissipation at all in these layers. 
This does not mean that in reality no dissipation exists in this layer, for instance, due to turbulence or liquid/solid friction at interfaces \citep[e.g.][]{2014ARA&A..52..171O, 2018haex.bookE..24M}. 
Nevertheless, such processes are not considered in our viscoelastic formalism, and therefore any dissipation is assumed negligible in liquid layers.


\begin{table*}[]
\centering
\begin{tabular}{|l|l|l|l|l|l|l|}
\hline
Planet 				& Mass (kg) 			& Radius (km)  	& Water mass fraction (\%)	& Fe/Si (\% wrt $\oplus$)  	\\ 
\hline
\hline
$0.5~\Mearth$ planet	& $2.98\times10^{24}$ 	& $5,376$ 	& 0 						& 50 	 				\\ 
	 				&  					& $5,214$ 	&   						& 100				\\ 
	 				&  					& $5,088$ 	&   						& 150 	 			\\ 
\hline
Earth-like 				& $5.95\times10^{24}$ 	& $6,573$ 	& 0 						& 50 	 				\\ 
	 				&  					& $6,364$ 	&   						& 100				\\ 
	 				&  					& $6,217$ 	&   						& 150 	 			\\ 
\hline
$5~\Mearth$ super Earth 	& $2.98\times10^{25}$	&  $10,246$	&  0						& 50		 			\\ 
				 	& 					&  $9,908$	&   						& 100	 			\\ 
				 	& 					&  $9,670$	&   						& 150	 			\\ 
\hline
$10~\Mearth$ super Earth	& $5.95\times10^{25}$ 	& $12,216$	&  0						& 50					\\ 
					& 				 	& $11,823$	&   						& 100				\\ 
					& 				 	& $11,547$	&   						& 150				\\ 
\hline
\hline
TRAPPIST-1e 			& $4.60\times10^{24}$	& $5,800$		&  0						&  138  				\\ 
 					&  					& $6,002$		&  5						&  150				\\
\hline
TRAPPIST-1f 			& $5.56\times10^{24}$	& $6,670$		&  11						&  100  				\\ 
 					&  					& $6,666$		&  14.5					&  150				\\
\hline
TRAPPIST-1g 			& $6.83\times10^{24}$	& $7,314$		&  19						&  100  				\\ 
 					&  					& $7,308$		&  22.5					&  150				\\
\hline
TRAPPIST-1h 			& $1.97\times10^{24}$	& $4,924$		&  10						&  100  				\\ 
 					&  					& $4,930$		&  13.5					&  150				\\
\hline
\hline
Venus 				& $4.85\times10^{24}$	& $6,152$		& 0						&  $\sim$100 			\\ 
\hline
\end{tabular}
\caption{Characteristics of the planets considered in this article. As we are not studying the impact of Earth's ocean, we are not treating Earth in itself but just an Earth-like planet of same mass and radius. However, we are treating Venus with as much details as we can based on the literature on this planet (following \citealt{2017JGRE..122.1338D}, see Section~\ref{Venus_structure}).} 
\label{table1}
\end{table*}

\subsubsection{Specific model for Venus}\label{Venus_structure}


We follow the work from \citet{2017JGRE..122.1338D} for the structural profiles of Venus. 
The mass and radius of Venus used in this model are given in Table~\ref{table1}.
We refer to this publication for further details for the interior model for Venus.
We use the temperature and pressure profiles from \citet{2010Icar..207..564S} (hereafter S10, cold profile) and \citet{2012JGRE..11712003A} (hereafter AT12, hot profile).


\begin{figure}     
\begin{center}
\includegraphics[width=1\linewidth]{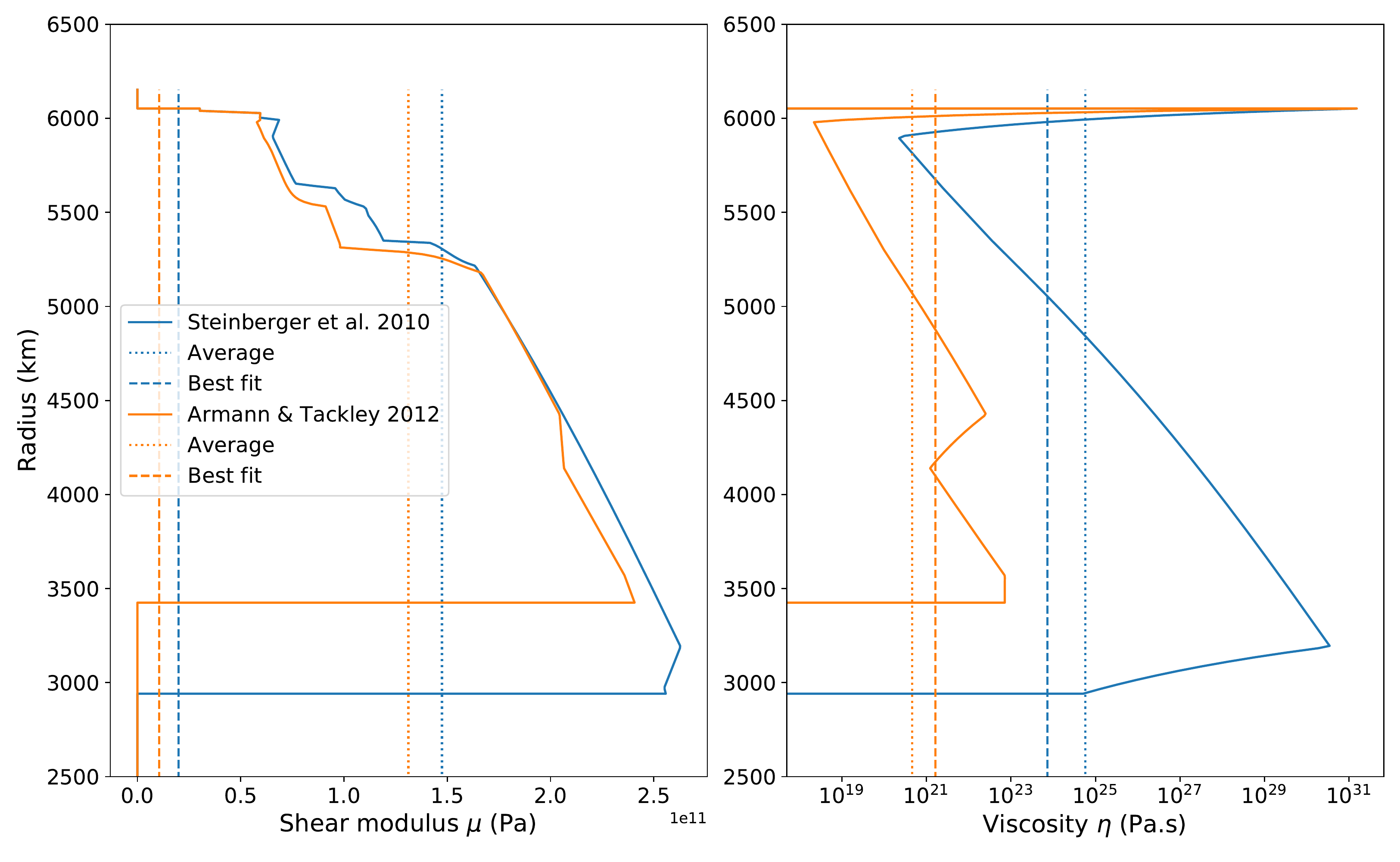}
\caption{Shear modulus and viscosity profiles for the two end-member interior models considered here. The viscosity $\eta$ is calculated as in \citet{2017JGRE..122.1338D}, using Eq.~\ref{viscosity_formula} of this work. The dotted lines represent the average of shear modulus and viscosity of the two profiles. The dashed lines correspond to the best fit of the dissipation of the multi-layer planet (see Section~\ref{Venus_dissip}).}
\label{plot_profile_mu_eta}
\end{center}
\end{figure} 

Radial density $\rho$ and seismic velocities $V_p$ and $V_s$ are computed from hydrostatic pressure, temperature, and composition using the Perple\_X program \citep{2005E&PSL.236..524C} developed by James Connolly (\url{http://www.perplex.ethz.ch}) in the mantle and from PREM extrapolation in the metallic core \citep{1981PEPI...25..297D}. 
The Perple\_X method computes phase equilibria and uses the thermodynamics of mantle minerals developed by \citet{2011GeoJI.184.1180S}.

As in the tests performed in \citet{2017JGRE..122.1338D}, we use the following formula to compute the viscosity as a function of the temperature and pressure profiles:
\begin{equation}\label{viscosity_formula}
\eta = \frac{1}{2}A_0^{-1}d^{2.5}\exp{\left(\frac{E_a + PV_a}{RT}\right)},
\end{equation}
where $E_a$, $V_a$ and $A_0$ are parameters depending on the material and d is the grain size.
Here, we consider dry olivine: $E_a = 300$~kJ.mol$^{-1}$, $V_a = 6$~cm$^3$.mol$^{-1}$, and $A_0 = 6.08 \times 10^{-19}$~Pa$^{-1}$.s$^{-1}$. 
We consider here a grain size d equal to 0.68~mm so that the factor $1/2~A_0^{-1}d^{2.5} = 10^{10}$~Pa.s.
The shear and bulk moduli are calculated from the density and the seismic velocity $V_s$ and $V_p$ as follows: $\mu = \rho V_s^2$ and $K=\rho V_p^2-4/3\mu$.
The corresponding shear modulus and the viscosity profiles are shown in Figure~\ref{plot_profile_mu_eta}.

The colder profile (S10) results in a much higher viscosity, and a shear modulus slightly larger than the hot profile (AT12). 
These two profiles can be considered as end members.
Figure~\ref{plot_profile_mu_eta} displays  the average parameter profiles for these two end-member interior models.
Here and in the following sections, the averages of shear modulus and viscosity are calculated as follows \citep{2017JGRE..122.1338D}:
\begin{equation}\label{log_average}
    <\mu> = \exp{\left[\frac{1}{V_{\rm shell}} \int_{V_{\rm shell}}{\ln{\mu}dv}\right]},
\end{equation}
where $V_{\rm shell}$ is the volume of the shell where viscosity and shear modulus are non-zero.


\section{Dissipation of multi-layer planets as a function of the excitation frequency}\label{dissipation_vs_frequency}

\subsection{Earth-like planet and super-Earths}
\label{Earth}

\begin{figure}[htbp]     
\begin{center}
\includegraphics[width=1\linewidth]{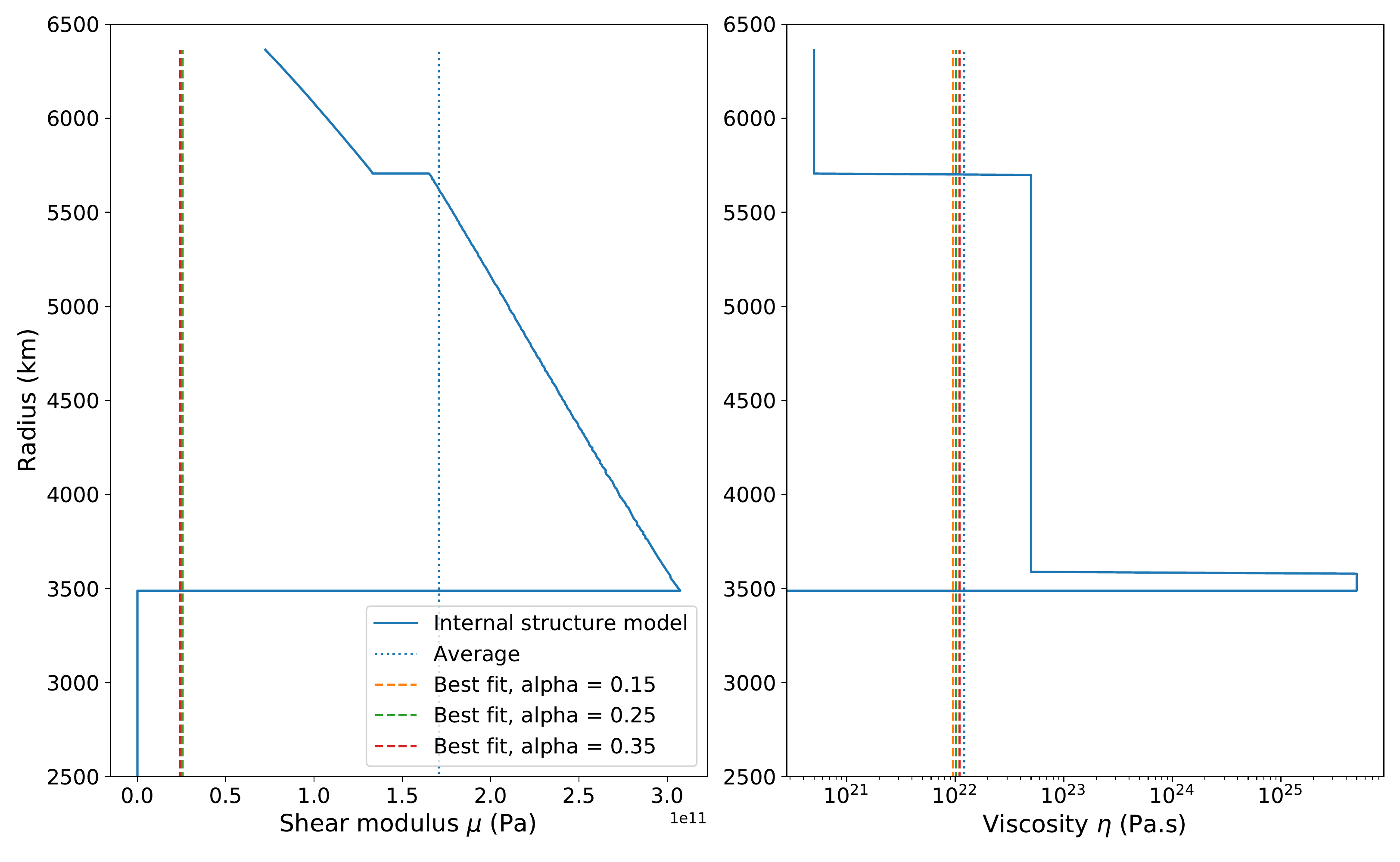}
\caption{Profiles of the shear modulus $\mu$ and the viscosity $\eta$ for an Earth-like planet (with Fe/Si = 100\%) in full lines. The dotted lines represent the average value of both quantities. The dashed lines represent the values obtained by fitting a homogeneous model to the dissipation response of the planet for 3 different values of the parameter $\alpha$ (see Figure~\ref{Fit_Earth_no_constraints_multilayer_vs_homogeneous_mean_fit}).}
\label{plot_profile_mu_eta_Earth_mean_fit}
\end{center}
\end{figure}

\begin{figure}[htbp]     
\begin{center}
\includegraphics[width=1\linewidth]{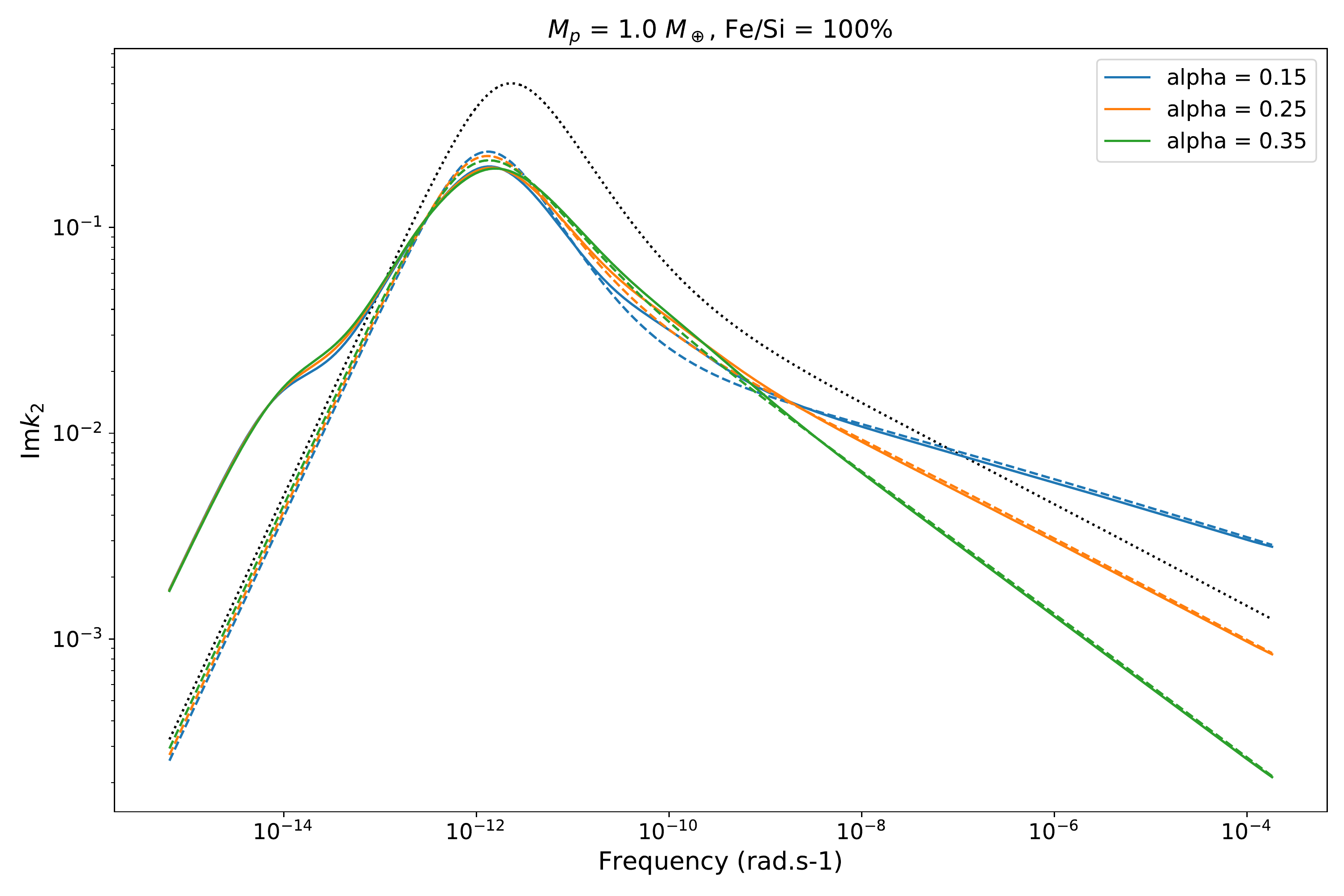}
\caption{Imaginary part of the Love number $\Imag k_2$ as a function of the excitation frequency for an Earth-like planet (without ocean). Full lines: Dissipation of a multi-layer body for three different $\alpha$ parameters (but $\alpha$ is the same for the different layers of these 3 cases). Dotted black line: Dissipation of a homogeneous body for which we assumed the averaged value of the multi-layer profiles for shear modulus and viscosity (see Fig.~\ref{plot_profile_mu_eta_Earth_mean_fit}). We also assumed a parameter $\alpha$ of 0.25 and assumed that the Maxwell Time is equal to the Andrade Time ($\zeta = 1$). Dashed lines: Best fit dissipation of a homogeneous body for the three different $\alpha$ (the fitted values of shear modulus and viscosity are plotted in Fig.~\ref{plot_profile_mu_eta_Earth_mean_fit} and all fit parameters including $\zeta$ parameter are listed in Table~\ref{table2}).}
\label{Fit_Earth_no_constraints_multilayer_vs_homogeneous_mean_fit}
\end{center}
\end{figure}

We use here a simple representation of Earth-like planets with 3 isoviscous layers and a shear modulus consistent with the PREM (as shown in Section~\ref{structure}, Fig.~\ref{plot_profile_mu_eta_Earth_mean_fit}). 
Figure~\ref{Fit_Earth_no_constraints_multilayer_vs_homogeneous_mean_fit} shows the imaginary part of the Love number, representing the amplitude of  dissipation, for  an ocean-less Earth-like planet as a function of the excitation frequency for a three-layer model and two homogeneous models. 

The full lines show the dissipation of the planet using the multi-layer model described in Section~\ref{multilayer_body}, using the profiles of Fig.\ref{plot_profile_mu_eta_Earth_mean_fit}.
The dotted line corresponds to the dissipation obtained considering a homogeneous planet following Section~\ref{Homogeneous_body} for which we assumed an averaged value for the shear modulus and the viscosity (Eq.~\ref{log_average}).
Finally the dashed lines correspond to the dissipation of a homogeneous planet for which we fitted the values of the shear modulus, the viscosity and $\zeta = \tau_A/\tau_M$, in order to obtain a dissipation amplitude similar to the three-layer model.

\begin{figure*}[htbp]     
\centering
\includegraphics[width=1\linewidth]{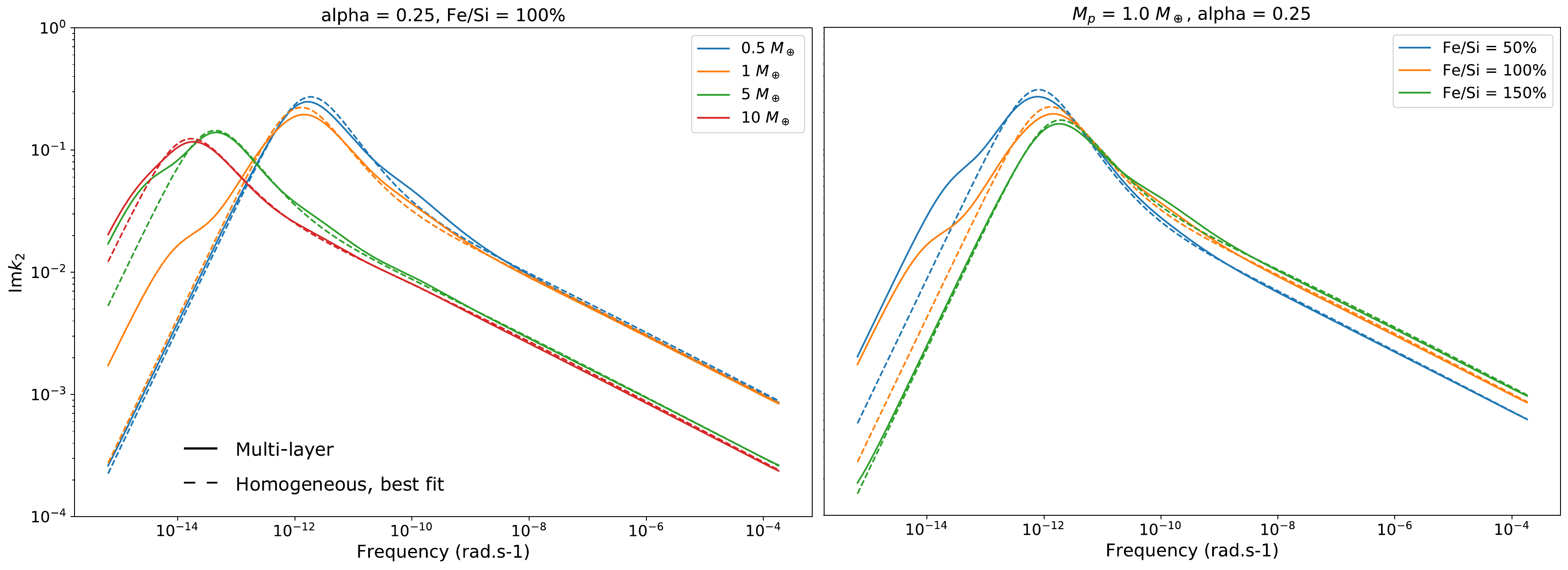}
\caption{As Fig.~\ref{Fit_Earth_no_constraints_multilayer_vs_homogeneous_mean_fit} but for different planetary masses and different Fe/Si ratios.}
\label{plot_comp_mass_iron}
\end{figure*} 

The fits were performed using the python library LMFIT (Non-Linear Least-Squares Minimization and Curve-Fitting for Python\footnote{It uses many of the optimization methods of scipy.optimize, see \url{https://lmfit.github.io/lmfit-py/index.html}}).
We used weights to force the fit to reproduce preferentially the amplitude and position of the dissipation maximum.
We evaluate the quality of our fits by calculating the Root Mean Square Error (RMSE) as follows
\begin{equation}\label{RMS}
\mathrm{RMSE} = \sqrt{\frac{1}{N}\sum_{i=1}^N \left( \Imag k_{2, {\rm multi}}(\omega_i) - \Imag k_{2, {\rm fit~homog}}(\omega_i) \right)^2},
\end{equation}
where $N$ is the number of individual excitation frequency values we consider, $\Imag k_{2, {\rm multi}}(\omega_i)$ is the imaginary part of the Love number we calculate using the multi-layer framework of Section~\ref{multilayer_body}, estimated at the excitation frequency $\omega_i$.
$\Imag k_{2, {\rm fit~homog}}(\omega_i)$ is the imaginary part of the Love number we calculate using the homogeneous model of Section~\ref{Homogeneous_body} with the fitted parameters for shear modulus, viscosity and $\zeta$.
In our fits, we do not treat $\alpha$ as a free parameter, but we use the same value as the one that was used for the different layers of the multi-layer model.
This ensures that the slope at high frequencies is accurately reproduced. 
As Fig.~\ref{Imk2_homog_parameters} shows, shear modulus, viscosity and $\zeta$ parameter influence the amplitude (through $\mu$) and position (through $\eta$) of the maximum of dissipation and the relative weight of the maximum over the high frequency behavior (through $\zeta$).
Table~\ref{table2} shows the best-fit parameters for all planets considered here.

One result shown in Fig.~\ref{Fit_Earth_no_constraints_multilayer_vs_homogeneous_mean_fit} is that considering a homogeneous body of averaged viscosity and shear modulus leads to a non-negligible overestimation of the dissipation.
If we compare the result for this averaged dissipation to the dissipation corresponding to the multi-layer body of same $\alpha$ (the green full line), it is more than 2.5 times higher and the difference increases with increasing frequency.
The position of the dissipation maximum is also slightly shifted to the higher frequencies.
These two major differences can be understood by comparing the order of magnitudes of the values of the averaged shear modulus and viscosity with the values of the fitted shear modulus and viscosity (Fig.~\ref{plot_profile_mu_eta_Earth_mean_fit}), which allow us to best reproduce the multi-layer dissipation. 
The averaged viscosity is very close to the best fit values, while the averaged shear modulus is much higher than the best fit values.
Fig.~\ref{Imk2_homog_parameters}c shows that at constant viscosity, the higher the shear modulus, the higher the peak of dissipation maximum and the more shifted the peak to the high frequencies, keeping the low frequency slope super-imposed.
Fig.~\ref{Fit_Earth_no_constraints_multilayer_vs_homogeneous_mean_fit} shows that exact behavior, which means that the dissipation using a homogeneous body with average values perform fairly well for the very low frequencies and is an overestimation at frequencies higher than the one corresponding to the maximum of dissipation.
Note that computing the logarithmic volume-weighted average as we did (Eq.~\ref{log_average}) yields a similar viscosity ($\sim 10^{22.1}$~Pa.s) as the best fits (around $10^{22.2}$~Pa.s, see Table~\ref{table2}), but that is not the case if we perform a linear volume-weighted average (which amounts to $\sim 10^{23.8}$~Pa.s) as is done in \citet{2018A&A...613A..37B}. 
The higher viscosity then leads to a shift of the peak to lower frequencies (see Fig.~\ref{Imk2_homog_parameters}b), which means that using a linear volume-weighted average leads to an underestimation of the dissipation over the whole frequency range.
The viscosity varies by several orders of magnitude between the low viscosity layers (upper mantle and bottom thermal boundary layer) and the high viscosity ones (top boundary layer and lower mantle). 
The linear average of the viscosity is therefore much more representative of the high viscosity layers. 
Performing a logarithmic average allows to compute a viscosity that better represents the low viscosity layers, without being too far from the high viscosity ones. 
We believe that, the viscosity being a key factor in controlling the position of the maximum of dissipation, the consideration of the low viscosity layers is important to better approximate a layered media using a constant parameter. 
This is confirmed by the fact that the best-fitting constant viscosity is very close to the logarithmic average of the varying viscosity.

Fig.~\ref{Fit_Earth_no_constraints_multilayer_vs_homogeneous_mean_fit} shows that the fit quality is better at high frequencies than at low frequencies where we see a bump occurring around $\omega\approx5\times10^{-13}$~rad.s$^{-1}$.
This low frequency behavior is due to the fact that the lower mantle has a higher viscosity value than the upper mantle. 
This leads to the secondary dissipation peak explaining the curve bump at low frequencies, around $10^{-14}$~rad.s$^{-1}$.


Figure~\ref{plot_comp_mass_iron} shows the dissipation curves but for planets with different masses (left panel) and Fe/Si ratios (right panel).
The higher the mass of the planet, the lower the dissipation peak. 
For an Earth-like planet, the maximum is about $\Imag k_2 \approx 0.2$, while it is $0.13-0.15$ for the 5~$\Mearth$ planet and $0.11$ for the 10~$\Mearth$ planet.
The higher the mass of the planet, the lower the frequency of the maximum of dissipation.
We provide the best-fit values of shear modulus and viscosity for different $\alpha$ and Fe/Si ratios in Table~\ref{table2}.

\begin{sidewaystable*}[htbp]
\begin{center}
\caption{Fit parameters for all planets considered here. The Maxwell Time $\tau_M$ is also given.}
\vspace{0.1cm}
\begin{tabular}{|c|c||c|c|c|c|c|c|c|c|c|c|c|c|c|}
\hline
 				& Parameters 		& \multicolumn{3} {|c|} {$1~\Mearth$ planet}	& \multicolumn{3} {|c|} {$5~\Mearth$ planet}	&  \multicolumn{3} {|c|} {$10~\Mearth$ planet}	&  	\multicolumn{2} {|c|} {TRAPPIST-1e}	&  	\multicolumn{2} {|c|} {Venus}\\
\hhline{~~-------------}
 				&  fit				& \multicolumn{3} {|c|} {Fe/Si}				& \multicolumn{3} {|c|} {Fe/Si}				& \multicolumn{3} {|c|} {Fe/Si}				&  \multicolumn{2} {|c|} {Fe/Si}			&  \multicolumn{2} {|c|} {Profile}	\\
 				&  				&  50\%		& 100\%		& 150\%		& 50\%		& 100\%		& 150\%		& 50\%		& 100\%		& 150\%		& 138\%		& 150\%		&  hot		& cold	\\
\hline
				& $\log\mu$ (Pa)	& 10.52		& 10.40		& 10.31		& 10.93 		& 10.76		& 10.63		& 11.18		& 11.00		& 10.86		& 10.25		& 9.97 		& 10.02		& 10.30	\\
				& $\log\eta$ (Pa~s)	& 22.27		& 22.01		& 21.79		& 23.93		& 23.93		& 23.92		& 24.66		& 24.59		& 24.55		& 21.76		& 21.64		& 21.21		& 23.86	\\
$\alpha = 0.25$	& $\zeta$				& 19.28		& 6.15		& 3.65		& 35.08		& 8.62		& 2.62		& 10.25		& 3.79		& 1.77		& 4.15		& 1.18		& 0.89	 	& 89.52 	\\
				& RMS			& 1.24e-2		& 7.62e-3		& 3.38e-3		& 1.12e-2		& 7.72e-3		& 5.66e-3		& 6.57e-3		& 3.24e-3		& 1.62e-3		& 2.90e-3		& 1.56e-2		& 2.16e-2		& 1.20e-2	\\	
				& $\tau_M$ (s)		& 5.62e11		& 4.07e11		& 3.02e11		& 1.00e13		& 1.48e13		& 1.95e13		& 3.02e13		& 3.89e13		& 4.90e13		& 3.24e11		& 4.68e11		& 1.55e11		& 3.63e13	\\
				& $\tau_M$ (day)	& 6.51e6    	& 4.72e6    	& 3.50e6    	& 1.16e8    	& 1.71e8    	& 2.26e8    	& 3.50e8    	& 4.50e8    	& 5.67e8    	& 3.75e6    	& 5.41e6    	& 1.79e6    	& 4.20e8	\\
\hline
				& $\log\mu$ (Pa)	& 10.53 		& 10.41		& 10.32		& 10.94 		& 10.77		& 10.63		& 11.19 		& 11.00		& 10.86		& 10.24		& 9.96		& 10.01		& 10.34	\\
				& $\log\eta$ (Pa~s)	& 22.24		& 21.98		& 21.78		& 23.91 		& 23.89		& 23.85		& 24.64		& 24.58		& 24.53		& 21.76		& 21.65		& 21.19		& 23.82	\\
$\alpha = 0.15$	& $\zeta$				& 282.08		& 46.94		& 17.84		& 204.18		& 33.40		& 7.83		& 43.12		& 11.09		& 4.23		& 23.19		& 6.13		& 4.21		& 4999.98\\
				& RMS			& 1.48e-2		& 9.19e-3		& 5.45e-3		& 1.20e-2		& 9.01e-3		& 7.37e-3		& 7.27e-3		& 3.97e-3		& 2.43e-3		& 4.05e-3		& 1.54e-2		& 2.38e-02	& 1.93e-2	\\	
				& $\tau_M$ (s)		& 5.13e11		& 3.72e11		& 2.88e11		& 9.33e12		& 1.32e13		& 1.66e13		& 2.82e13		& 3.80e13		& 4.68e13		& 3.31e11		& 4.90e11		& 1.51e11		& 3.02e13 \\
				& $\tau_M$ (day)	& 5.94e6    	& 4.30e6    	& 3.34e6    	& 1.08e8    	& 1.53e8    	& 1.92e8    	& 3.26e8   	& 4.40e8    	& 5.41e8    	& 3.83e6    	& 5.67e6    	& 1.75e6    	& 3.50e8	\\
\hline
				& $\log\mu$ (Pa)	& 10.51		& 10.39		& 10.30		& 10.92 		& 10.76		& 10.64		& 11.18		& 11.00		& 10.86		& 10.25		& 10.01 		& 10.04		& 10.26	\\
				& $\log\eta$ (Pa~s)	& 22.31		& 22.04		& 21.80		& 23.96		& 24.00		& 24.07		& 24.67		& 24.63		& 24.61		& 21.77		& 21.66		& 21.22		& 23.91	\\
$\alpha = 0.35$	& $\zeta$				& 4.98		& 2.22		& 1.67		& 11.90		& 3.05		& 0.90		& 4.90		& 1.95		& 0.91		& 1.81		& 0.36		& 0.38		& 6.73 	\\
				& RMS			& 1.05e-2		& 6.39e-3		& 1.73e-3		& 1.03e-2		& 6.48e-3		& 4.05e-3		& 5.80e-3		& 2.55e-3		& 1.36e-3		& 1.64e-3		& 1.60e-2		& 1.78e-2		& 4.50e-3	\\	
				& $\tau_M$ (s)		& 6.31e11		& 4.47e11		& 3.16e11		& 1.10e13		& 1.74e13		& 2.69e13		& 3.09e13		& 4.27e13		& 5.62e13		& 3.31e11		& 4.47e11		& 1.51e11		& 4.47e13	\\
				& $\tau_M$ (day)	& 7.30e6    	& 5.17e6    	& 3.66e6    	& 1.27e8    	& 2.01e8    	& 3.12e8    	& 3.58e8    	& 4.94e8    	& 6.51e8    	& 3.83e6    	& 5.17e6    	& 1.75e6    	& 5.17e8	\\
\hline
\end{tabular} 
\tablefoot{We remind the reader that: $\zeta = \tau_A/\tau_M$. The fitted values for TRAPPIST-1e are given for the case where the viscosity of the high pressure ice is  $10^{16}$~Pa.s, which corresponds to Fig.~\ref{Comp_Trappist1e_highlow_iron}.}
\label{table2} 
\end{center}
\end{sidewaystable*}

\subsection{Venus}
\label{Venus_dissip}

\begin{figure}[htbp]     
\begin{center}
\includegraphics[width=1\linewidth]{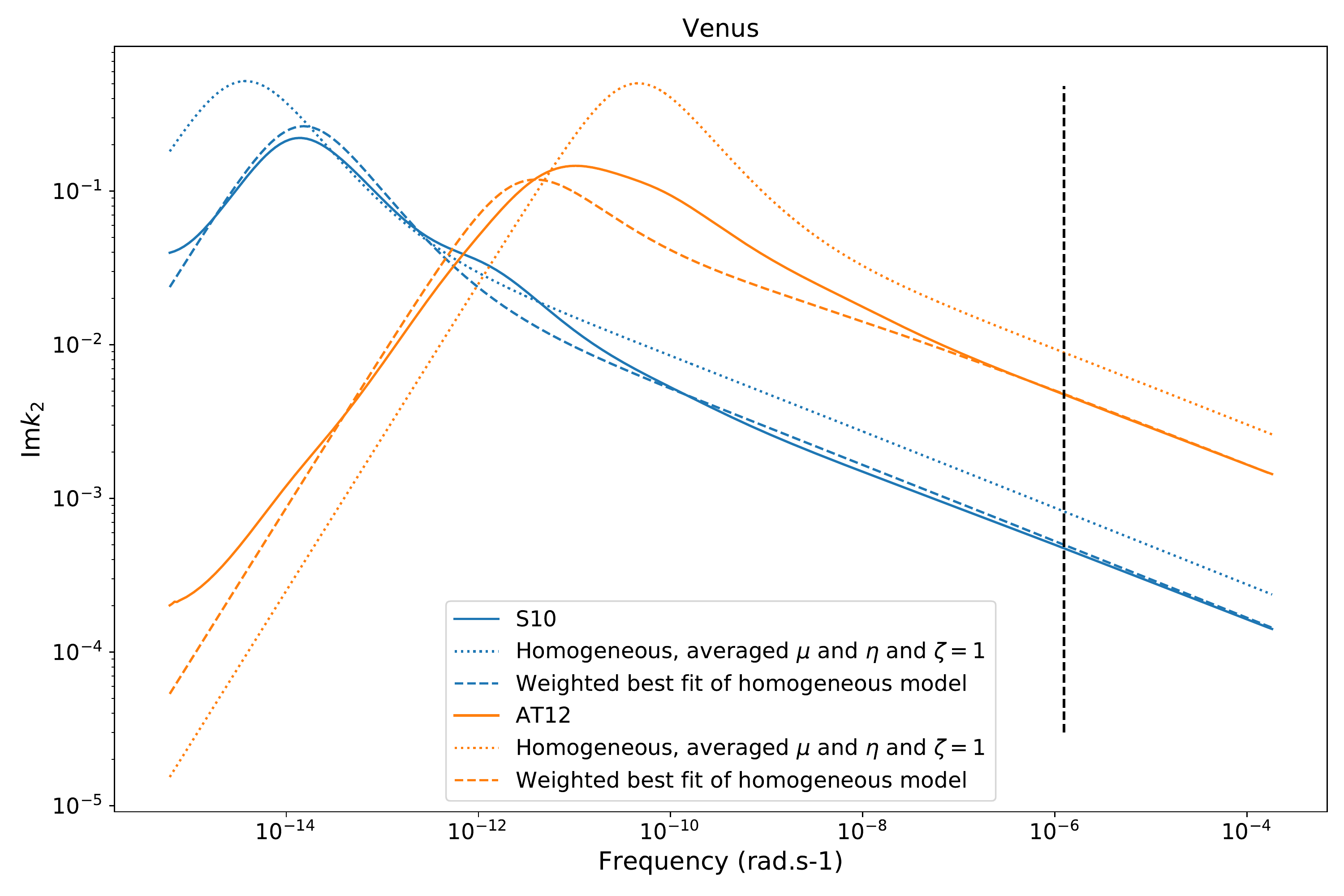}
\caption{Imaginary part of the Love number $\Imag k_2$ as a function of the excitation frequency assuming $\alpha = 0.25$. Full lines: $\Imag k_2$ as calculated by the profiles of \citet{2017JGRE..122.1338D} for the two different models (hot: orange, cold: blue). Dashed lines: Best fit of $\Imag k_2$ for a homogeneous body. Dotted lines: Obtained with the averaged values of $\mu$ and $\eta$ from Fig.~\ref{plot_profile_mu_eta}. The black vertical dashed line represents the excitation frequency of Venus ($\omega_{{\rm Venus}} = 2(n_{{\rm Venus}}-\Omega_{{\rm Venus}})$).}
\label{Fit_Venus_no_constraints_multilayer_vs_homogeneous}
\end{center}
\end{figure}

Figure~\ref{Fit_Venus_no_constraints_multilayer_vs_homogeneous} shows the frequency dependency of the dissipation curves for the multi-layer models of Venus (full lines), for a homogeneous body with averaged values of shear modulus and viscosity (dotted lines), and for a homogeneous body with best fit values of shear modulus and viscosity for the two end-member (cold/hot) internal structure models. 

\begin{figure*}[htbp]     
\centering
\includegraphics[width=1.0\linewidth]{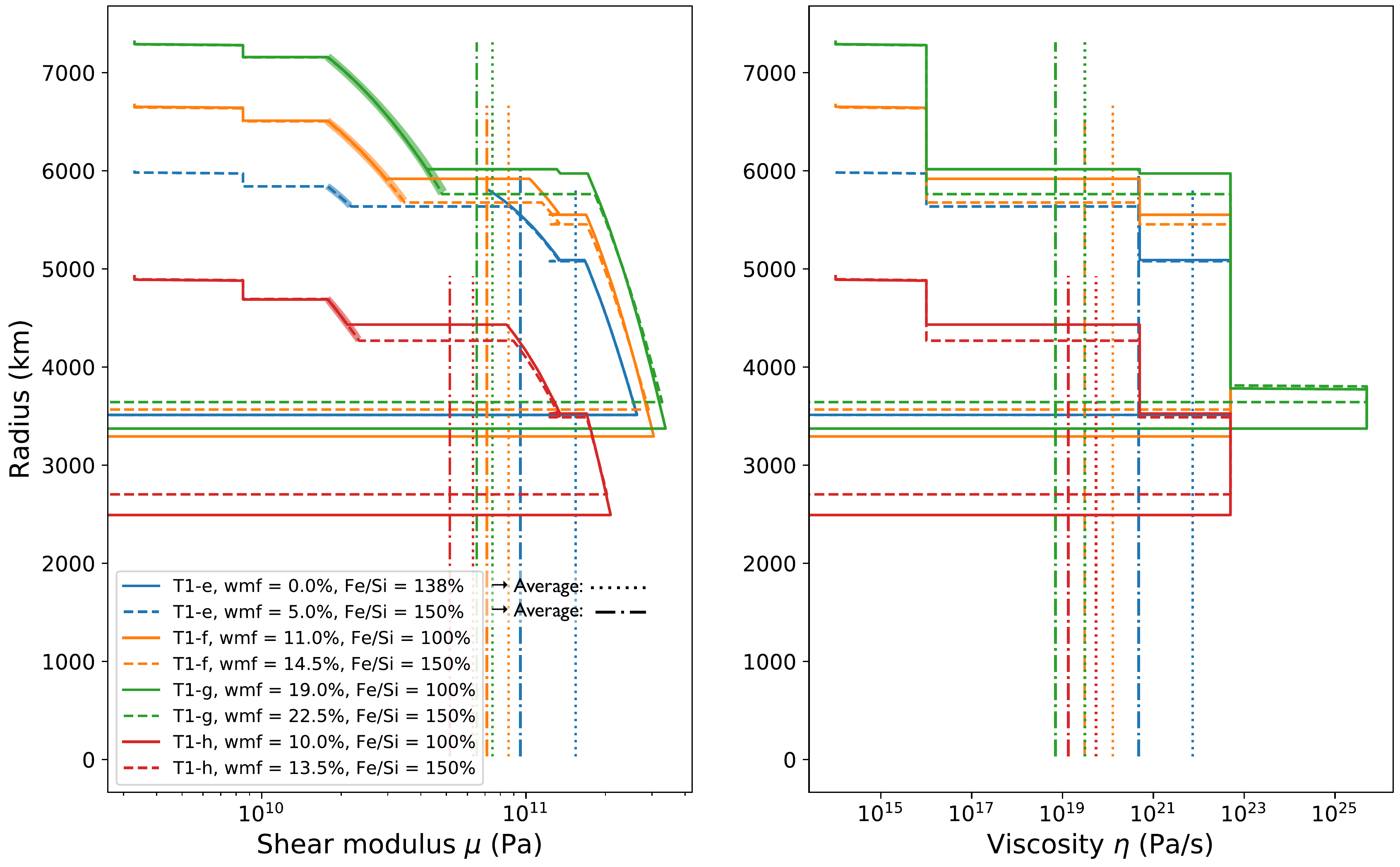}
\caption{Shear modulus and viscosity profiles for the outer TRAPPIST-1 planets for the highest iron content and therefore the highest ice content. \textit{wmf} means water mass fraction, and the ratio Fe/Si is given respect to the Earth. On the left panel, the high pressure ices layer can be identified as a bolder part of the line.}
\label{plot_profile_mu_eta_T1_outer_planets}
\end{figure*} 

The two internal structure models lead to quite different characteristics: the ``cold'' profile (S10) leads to a slightly higher dissipation maximum than the ``hot'' profile (AT12) and the maximum dissipation occurs at shorter frequencies.
This behavior can be understood from the profiles of shear modulus and viscosity for the two different structures (see Figs.~\ref{plot_profile_mu_eta} and \ref{Imk2_homog_parameters}): both quantities are higher for the cold profile than for the hot profile and this leads to a higher dissipation maximum and a maximum shifted towards shorter frequencies.  
Both models lead to very different dissipation values at the present-day tidal frequency of Venus (indicated with the black dashed line). 
The difference in dissipation rate between the cold and hot profiles suggests that a fast-rotating early Venus would despin about ten times more quickly to its present-day rotation state for the hot interior  scenario than for the cold one. 
This indicates that the past rotation evolution of Venus is rather sensitive to the thermal evolution of its mantle, and that detailed thermal modeling is required to accurately reconstruct its rotation history \citep{2018EPSC...12..856D}.  

Considering a homogeneous body of averaged shear modulus and viscosity leads to an overestimation of the dissipation in the high frequency regime. 
The ``hot'' profile case shows a higher overestimation visible around the dissipation maximum: it is higher and slightly shifted to the higher frequencies. 
For the ``cold'' profile, the dissipation maximum is also higher but shifted to the lower frequencies.
This behavior can be understood when comparing the averaged value of shear modulus and viscosity with the best fit values (see Fig.~\ref{plot_profile_mu_eta}) and Fig.~\ref{Imk2_homog_parameters}.
Both cases have a higher averaged shear modulus than the best fit value which leads to a higher dissipation maximum, slightly shifted to the higher frequencies.
For the ``hot'' profile, the averaged viscosity is lower than the best fit viscosity, which contributes to shifting the dissipation maximum to even higher frequencies.
For the ``cold'' profile, the averaged viscosity is higher than the best fit viscosity, which shifts the dissipation maximum to lower frequencies and overcomes the shear modulus-driven shift.
Like for super-Earths in the previous section, the fits are however not perfect. 
The dissipation of the multi-layer structure leads to a secondary peak in the dissipation around a frequency of $10^{-12}$~rad.s$^{-1}$ for the ``cold'' profile in blue and very close to the main peak for the ``hot'' profile around a frequency of $10^{-10}$~rad.s$^{-1}$.
Considering the dissipation of a homogeneous body leads therefore here to a small under-estimation of the dissipation. 
However, as these small differences occur at very low frequencies, their impacts on the rotation evolution should be limited. 
The detailed consequences in term of spin and eccentricity evolution will be investigated in a follow-up article. 
The fitted values of the shear modulus and viscosity can be found in Table~\ref{table2} for $\alpha = 0.25$ (used for Fig.~\ref{Fit_Venus_no_constraints_multilayer_vs_homogeneous}) as well as the two other values of $\alpha$ considered here.

\subsection{TRAPPIST-1e}

The two internal structure models we use here for TRAPPIST-1e differ by the Fe/Si ratio and the consequent presence of an ice layer (see Table~\ref{table1}).
Contrary to all the models we have discussed so far, we will investigate here the effect of the presence of an ice layer on the dissipation of a terrestrial planet. 

Figure~\ref{plot_profile_mu_eta_T1_outer_planets} also shows the profiles for two different possible internal structures of TRAPPIST-1e: a dense ice-less structure and a less dense icy structure. 
The ice-less structure is very similar to the one of the Earth-like planet with only three layers. 
Note that the value of Fe/Si ratio has to be higher than that of the Earth in order to reproduce the observed radius (138\%).
The second structure has a $\sim400$~km ice layer at the surface, where the three different phases of water ice (Ice I, Ice VI, Ice VII) are present.
The amount of ice for this structure represents only 5\% of the mass of the planet.
As ice is much less dense than rocks, an even higher iron content (150\%) is needed to reproduce the observed radius.
The properties of these two different cases are given in Table~\ref{table1}.

\begin{figure}[htbp]     
\begin{center}
\includegraphics[width=1.0\linewidth]{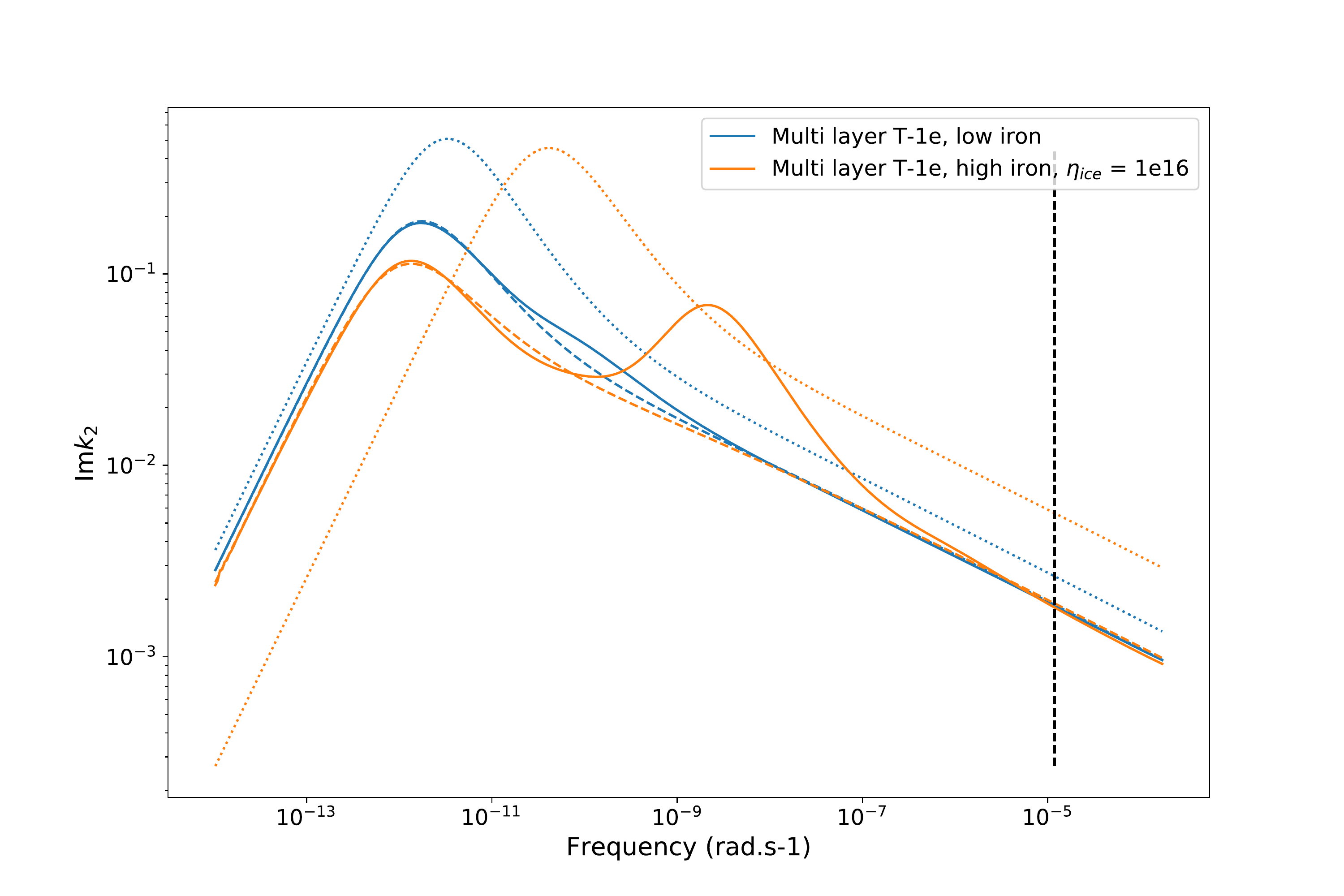}
\caption{Imaginary part of the Love number $\Imag k_2$ as a function of the excitation frequency for two different internal structures of TRAPPIST-1e, assuming $\alpha = 0.25$. The iron-poor structure (ice-less structure) is in blue and the iron-rich structure (with an ice layer) is in orange. Full lines: multi-layer model. Dashed lines: the corresponding best fits of a homogeneous model. Dotted lines: homogeneous model with averaged viscosity and shear modulus. The black vertical line represents the excitation frequency of TRAPPIST-1e, which is the orbital frequency in the following hypotheses: the rotation is synchronized, the obliquity is zero and the eccentricity is small.}
\label{Comp_Trappist1e_highlow_iron}
\end{center}
\end{figure}

Figure \ref{Comp_Trappist1e_highlow_iron} shows the frequency dependence of the dissipation for these two different structures. 
The ice-poor structure (low iron content) is very similar to the super-Earths of Section~\ref{Earth} and Venus (Section~\ref{Venus_dissip}), resulting in similar fit performance.
The values of the fit are given in Table~\ref{table2} for all values of $\alpha$.
Table~\ref{table2} also shows the corresponding values of the Maxwell time for the different fits. 
The obtained values of the Maxwell time are orders of magnitude higher than the values of $10^{-2}-10^{-1}$~days taken by \citet{2018ApJ...857..142M} for TRAPPIST-1 planets.
This means that the excitation frequency corresponding to the maximum dissipation is shifted to higher frequencies than in our work and that the maximum dissipation is also much higher in their case than in ours (see Fig.~\ref{Imk2_homog_parameters}).
We therefore expect the dissipation they calculate to be strongly overestimated compared to the one we compute in the next Section~\ref{tidal_heating_T1}.

The behavior of $\Imag k_2$ with frequency is very different when there is an ice-layer.
Figure~\ref{Comp_Trappist1e_highlow_iron} shows two peaks occurring at different frequencies.
The shorter frequency corresponds to the frequency of the rocky part of the planet (it is indeed very similar to the excitation frequency at which the maximum of the response occurs for the ice-less structure). 
The second peak occurring at larger frequencies corresponds to the behavior of the ice.
For this specific case here, we use a constant viscosity of $10^{16}$~Pa.s for the high pressure ice layer as shown on Figure~\ref{plot_profile_mu_eta_T1_outer_planets}.
This double-peaked feature makes the quality of the fit very poor. 
 
A multilayer planet with different rocky layers can be quite well approximated by a homogeneous body with fitted values for shear modulus and viscosity but if the planet has an ice layer, then this approximation is no longer valid.
By construction, the homogeneous body does not allow to reproduce the double-peaked feature of the dissipation\footnote{Note that adding two homogeneous models with different weights to reproduce the two peaks does not work either in so far as it does not reproduce the high frequency behavior.}, so for icy bodies it is necessary to take into account the real multi-layer structure to correctly estimate the dissipation \citep[or try fitting with a different model, e.g.][but this is out of the scope of this article]{2018ApJ...857...98R}.
The homogeneous body approximation particularly fails on a range of frequencies that depends on the viscosity of the ice layer considered.

As the viscosity of high pressure ice is relatively unknown, we tested two other values of viscosity:  $10^{14}$~Pa.s \citep[close to the value of $10^{13}$~Pa.s obtained from laboratory experiments given in][]{1981Natur.292..225P} and $10^{18}$~Pa.s.
This range is roughly what was considered in \citet{2018Icar..299..133K} for ice layers, which come from the two preceding articles, \citet{1989PEPI...55...10S} and \citet{2001AREPS..29..295D}.

\begin{figure}[htbp]     
\begin{center}
\includegraphics[width=1.0\linewidth]{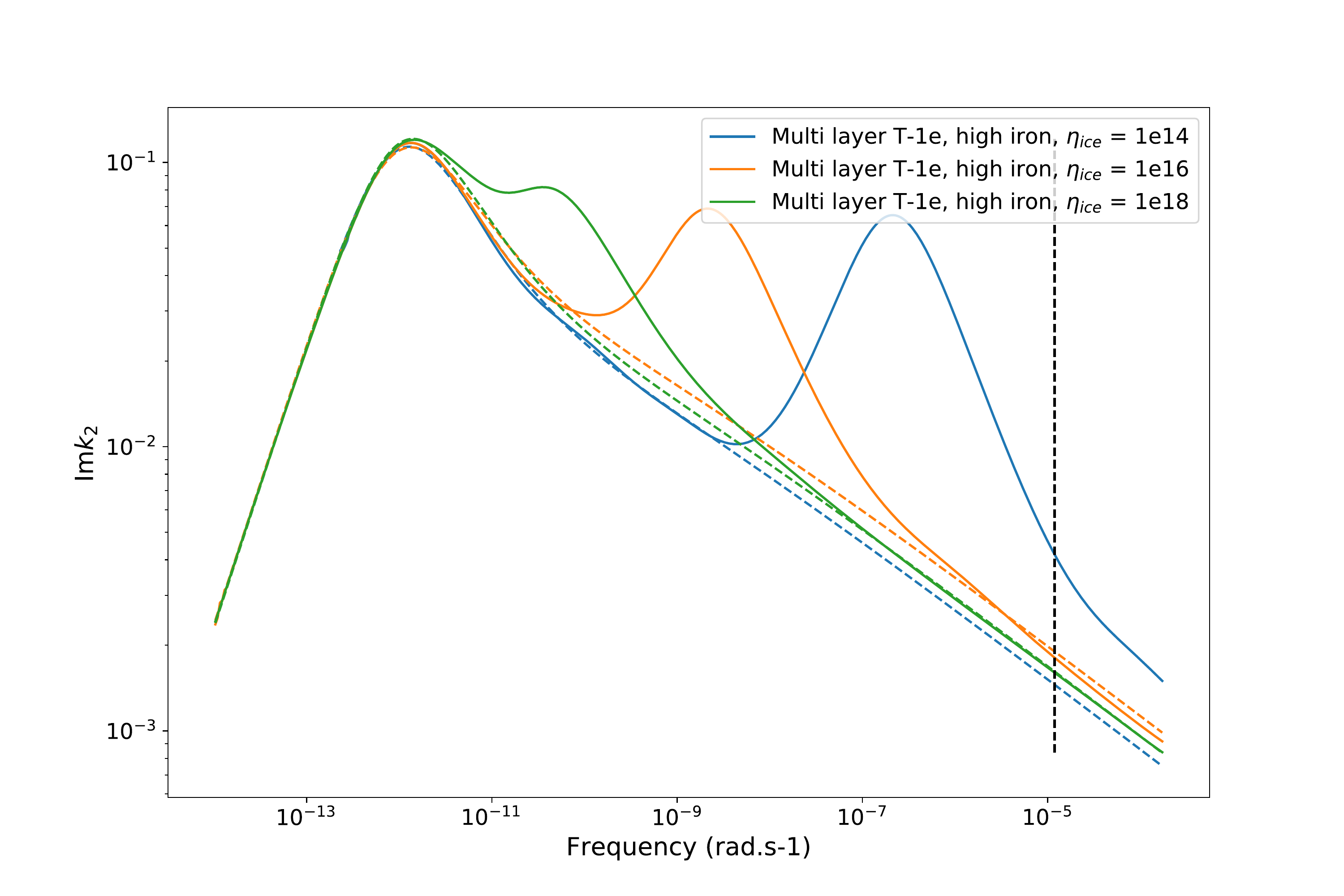}
\caption{Same as Fig.~\ref{Comp_Trappist1e_highlow_iron} but for different viscosities of the ice layer for the iron-rich internal structure for TRAPPIST-1e, assuming $\alpha = 0.25$.} 
\label{Comp_Trappist1e_high_iron_ice}
\end{center}
\end{figure}

Figure~\ref{Comp_Trappist1e_high_iron_ice} shows the behavior of the dissipation response when the viscosity of the high pressure ice layer varies from $10^{14}$~Pa.s to $10^{18}$~Pa.s.
When the viscosity of the ice layer increases, the peak linked to the dissipation in the ice layer shifts towards the shorter frequencies, which is compatible with the behavior that we would expect for a homogeneous body (see Figure~\ref{Imk2_homog_parameters}).
The maximum of the second peak decreases slightly when the viscosity of the ice layer decreases but as the peak shifts to higher frequencies, the planet becomes significantly more dissipative for a wider range of frequencies.
Given the poor quality of the fit, we do not give the fitted values of shear modulus and viscosity for the two ice-layer viscosities  $10^{14}$~Pa.s and $10^{18}$~Pa.s.
For the following, we assume a viscosity of  $10^{16}$~Pa.s for the high pressure ice layer of the TRAPPIST-1 planets.

\section{Tidal heating of the TRAPPIST-1 planets}\label{tidal_heating_T1}

We derive here the dissipation profile within the TRAPPIST-1 planets using the multi-layer structure approach.
Given the radius \citep{2017Natur.542..456G} and estimated masses \citep{2018A&A...613A..68G} of the TRAPPIST-1 planets, the density of the planets of TRAPPIST-1 is compatible with a large amount of volatiles \citep{2018A&A...613A..68G, 2018ApJ...865...20D}.
We focus here on the outer planets of the TRAPPIST-1 systems for which the volatiles are taken to be water ice.
We will study the inner planets in a following article, due to the additional difficulty that the volatiles (that we consider being water here) are in the fluid form in the external envelope.
Thus investigating those planets would require coupling the internal structure model to a model of an steam atmosphere in runaway in equilibrium with supercritical water fluid and possibly a magma ocean, which is out of the scope of this work.

In agreement with previous studies \citep{2017NatAs...1E.129L,2018A&A...612A..86T, 2018A&A...613A..68G}, we consider here that the planets are in synchronous rotation, have zero obliquities and have small eccentricities. 
We consider two cases for the eccentricities of the TRAPPIST-1 planets: the ones given by the TTV analysis of the system in \citet{2018A&A...613A..68G} and the smaller ones given in \citet{2018A&A...612A..86T}.
The lower eccentricities given in \citet{2018A&A...612A..86T} were obtained performing a N-body simulation of the system from \citet{2018A&A...613A..68G} taking into account the tidal forces and torques using a CTL equilibrium tide model \citep{2015A&A...583A.116B}.
The eccentricities for both cases are given in Table~\ref{table3}.

The planets' masses and radii for different water mass fractions (or iron content) are given in Table~\ref{table1}.
Figure~\ref{plot_profile_mu_eta_T1_outer_planets} shows the radial profiles for the shear modulus and viscosity of each of the four outer TRAPPIST-1 planets, derived from the density profile estimated from the planets' masses and radii following the same approach as in \citet{2019A&A...630A..70T}. 

In order to calculate the tidal heating profile in the TRAPPIST-1 planets, we use Equation 37 of \citet{2005Icar..177..534T}, which is valid for synchronous planets with small eccentricities
\begin{equation}\label{htide}
h_{\rm tide}(r) = -\frac{21}{10} n^5 \frac{\Rp^4 e^2}{r^2} H_\mu \Imag \tilde\mu, 
\end{equation}
where $n$ is the orbital frequency, $e$ is the eccentricity of the orbit, $r$ the radius at which the volumetric tidal heating is estimated.
$H_\mu$ represents the radial sensitivity to the shear modulus $\mu$ and is a real quantity.
It depends on the radial structure of the planet and on the $y_i$ functions introduced in Section~\ref{multilayer_body} as follows (from Eq. 33 of \citealt{2005Icar..177..534T} for $l = 2$)
\begin{equation}\label{Hmu}
\begin{split}
H_\mu 	&= \frac{4}{3}\frac{r^2}{|\tilde\kappa + 4/3\tilde\mu|^2} \left| y_2 - \frac{\kappa - 2/3\tilde\mu}{r}(2 y_1-6 y_3)\right|^2\\
 		&\quad -\frac{4}{3} r {\rm Re}\left\{ \frac{\dd y_1^*}{\dd r} (2y_1 - 6 y_3) \right\}\\
		&\quad + \frac{1}{3} \left| 2y_1 - 6y_3\right|^2  + 6 r^2|y_4^2|/|\tilde\mu|^2 + 24 |y_3|^2.
\end{split}
\end{equation}
$\Imag \tilde\mu$ in Eq. \ref{htide} is the imaginary part of the complex shear modulus given in Eq.~\ref{Hooke_general_mu}. 
As in \citet{2005Icar..177..534T}, we neglect the imaginary part of the complex incompressibility $\tilde\kappa$ thus assuming that no bulk dissipation occurs and that all dissipation is associated to shear deformation. 
In this formalism, $\Imag \tilde\mu$ contains all the information about the dissipation.

For comparison, we also calculate the volumetric heating of the Earth and Venus. 
These planets are not tidally locked but can be considered on coplanar, circular orbits. 
In that case, the expression of $h_{\rm tide}(r)$ is given by
\begin{equation}\label{htide_EV}
h_{\rm tide}(r) = \frac{3}{10} \Omega \frac{\G^2M_0^2\Rp^4}{a^6r^2} H_\mu \Imag \tilde\mu, 
\end{equation}
where $\Omega$ is the spin of the planet, $M_0$ and $a$ is respectively the mass of the Sun and the semi-major axis of Venus for the case of Venus and the mass and semi-major axis of the Moon in the case of the Earth.
For the case of Venus $\Omega = -2.99\times10^{-7}$~rad.s$^{-1}$ (corresponding to a rotation period of -243 day) and the excitation frequency is $\omega = 2(n-\Omega) = 1.24\times10^{-6}$~rad.s$^{-1}$ (excitation period of $\sim 56$~day).
For Earth, $\Omega = 7.27\times10^{-5}$~rad.s$^{-1}$ and the excitation frequency is $\omega = -1.41\times10^{-4}$~rad.s$^{-1}$ (period of $\sim -0.5$~day).

\begin{figure}[htbp]     
\begin{center}
\includegraphics[width=1.0\linewidth]{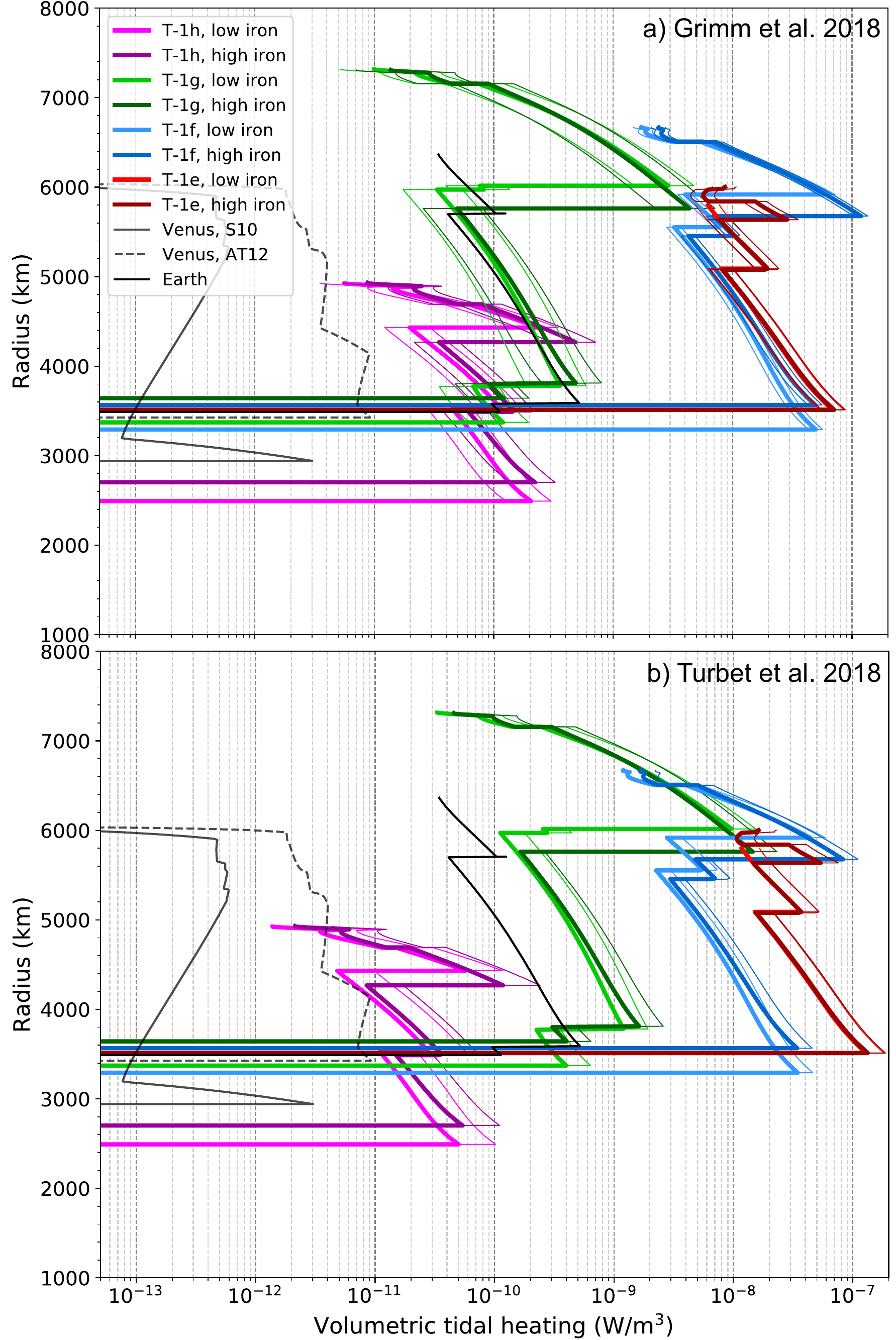}
\caption{Tidal heating profile of the outer TRAPPIST-1 planets for different eccentricities sets. Top panel: The eccentricities come from the TTV analysis of \citet{2018A&A...613A..68G}, the tidal heating profile was computed using the median eccentricity (thick line) and the standard deviation (thin lines) of the eccentricity are represented. Bottom panel: The eccentricities come from N-body simulations with tides, presented in \citet{2018A&A...612A..86T}. For comparison, the heating profiles of the solid Earth (black line) and Venus (grey lines: full and dashed for the ``cold'' and ``hot'' profiles respectively) were added.}
\label{plot_heat_profile_outer_T1}
\end{center}
\end{figure}

Figure~\ref{plot_heat_profile_outer_T1} shows the tidal heating profiles for the outer planets of TRAPPIST-1 for the different hypotheses in eccentricity and iron content, as well as the tidal heating profile of the rocky part of the Earth (black line) and of Venus for both ``cold'' (full grey line) and ``hot'' (dashed grey line) profiles. 
Panel a) of Fig.~\ref{plot_heat_profile_outer_T1} shows the profiles for the eccentricities of \citet{2018A&A...613A..68G} and panel b) shows the profiles for the eccentricities of \citet{2018A&A...612A..86T}. 
Within a given layer, the volumetric dissipation increases with depth, with huge variations between layers. 
The high pressure ice layer is the one which is responsible for the highest tidal heating, as this layer has a much lower viscosity ($\eta = 10^{16}$~Pa.s) than the mantle ($\eta > 10^{21}$~Pa.s). 
For all planets, we obtained on enhancement of tidal dissipation at the transition between the silicate mantle and the high-pressure ice mantle. 
We observe also that the enhancement of dissipation in the high pressure mantle is more pronounced with increasing ice fraction.  
For planet TRAPPIST-1g, which is the planet with the largest ice fraction, the tidal heating increases by about two order of magnitudes at the rock/ice interface. 

One of the most striking result is that planet f dissipates more energy than the closer-in planet e.
This is due to the the fact that its eccentricity is about twice as high as the eccentricity of planet e (see Table~\ref{table3}) and the fact that planet f has a much thicker ice mantle and a slightly larger radius than planet e, which is the densest of the four outer TRAPPIST-1 planets.

The change of tidal heating with depth is  mainly controlled by the behavior of the quantity $H_\mu \Imag \tilde\mu$. 
The transition between the rock mantle and the high-pressure ice layer is characterized by a drop in shear modulus $\mu$. 
The reduction of shear modulus in the high-pressure ice mantle relative to the rock mantle leads to an abrupt increase in  tidal flexing, resulting in an enhancement of shear deformation, described by the sensitivity parameter to shear deformation, $H_\mu$.  
Both quantities $H_\mu$ and $\Imag \tilde\mu$ increase towards the interior of the planet in the high pressure ice layer and are maximized at the interface.
$\Imag \tilde\mu$ depends on the quantities $\eta\omega$ vs $\mu^2$.
In our model, the viscosity $\eta$ of each layer is constant, while the shear modulus of the high pressure ices layer increases towards the interior of the planet (see Fig.~\ref{plot_profile_mu_eta_T1_outer_planets}, the bolder part of the curve corresponds to the high pressure ices layer).
Consequently, if the quantity $\eta\omega$ is much higher than $\mu^2$, then $\Imag \tilde\mu$ varies as $\mu^2$, which is the case for most of the layers of planet g and of all the other planets.

The difference of composition of the planets do not entail a radical difference in the dissipated energy, but is still statistically significant. 
Due to fact that most of the dissipation occurs in the high pressure ice layer, the planets with the higher iron content, which also have the higher quantity of water in order to fit the M-R relationship, have a higher dissipation.
The difference between the iron poor and iron rich compositions is the lowest for planet e, where the total energy dissipated differs by about 15\%.
This difference is of about 35\% for planets f and g and reaches 36\% for planet g.

\begin{table*}[]
\centering
\begin{tabular}{|l|l||l|l||l|l||l|l|}
\hline
\multicolumn{2}{|l||}{}			    & \multicolumn{2}{|c||}{Multi-layer model}   & \multicolumn{2}{|c||}{Homogeneous model}  & \multicolumn{2}{|c|}{Layer-averaged model}  \\ 
\hline
Planet 		& Eccentricity 			& Dissipation  	& Tidal heat flux 	        & Dissipation  	& Tidal heat flux	        & Dissipation  	& Tidal heat flux  	\\ 
			&					    & (TW)		    & (W.m$^{-2}$)		        & (TW)		    & (W.m$^{-2}$)	            & (TW)		    & (W.m$^{-2}$)	\\
\hline
\hline
T-1e 		& $4.52\times 10^{-3}$	& 9.3e0		    &  2.2e-2	                & 6.4e3         & 1.5e1	                    & 9.5e0         & 2.3e-2   \\ 
 			& $5.10\times 10^{-3}$ 	& 1.2e1		    &  2.8e-2	                & 8.2e3         & 1.9e1	                    & 1.2e1         & 2.9e-2	\\
			& $5.68\times 10^{-3}$	& 1.5e1		    &  3.5e-2	                & 1.0e4         & 2.4e1	                    & 1.5e1         & 3.6e-2	\\
\hline  
T-1f 		& $9.39\times 10^{-3}$	& 1.4e1		    &  2.5e-2	                & 6.5e3         & 1.3e-1	                & 1.5e1         & 2.6e-2	\\ 
 			& $1.01\times 10^{-2}$	& 1.6e1		    &  2.9e-2	                & 7.5e3         & 2.4e-1	                & 1.7e1         & 3.0e-2	\\
			& $1.08\times 10^{-2}$	& 1.8e1		    &  3.3e-2	                & 8.6e3         & 4.0e-1	                & 1.9e1         & 3.4e-2	\\
\hline
T-1g 		& $1.50\times 10^{-3}$ 	& 3.3e-1	    &  5.0e-4 	                & 8.6e1         & 1.3e-1	                & 3.6e-1        & 5.3e-4	\\ 
 			& $2.08\times 10^{-3}$	& 6.4e-1		&  9.6e-4	                & 1.6e2         & 2.4e-1	                & 6.9e-1        & 1.0e-3	\\
			& $2.66\times 10^{-3}$	& 1.1e0		    &  1.6e-3	                & 2.7e2         & 4.0e-1	                & 1.1e0         & 1.7e-3	\\
\hline
T-1h 		& $4.46\times 10^{-3}$	& 1.9e-2		&  6.3e-5 	                & 6.6e0         & 2.2e-2	                & 2.0e-2        & 6.4e-5	\\ 
 			& $5.67\times 10^{-3}$	& 3.1e-2		&  1.0e-4	                & 1.1e1         & 3.5e-2	                & 3.2e-2        & 1.0e-4	\\
			& $6.88\times 10^{-3}$	& 4.6e-2		&  1.5e-4	                & 1.6e1         & 5.2e-2	                & 4.7e-2        & 1.5e-4	\\
\hline
\end{tabular}
\caption{Range of possible eccentricities for the four outer planets of TRAPPIST-1 from \citet{2018A&A...613A..68G}, the corresponding dissipated energy and tidal heat fluxes for the low iron scenario for a multi-layer planet and for the homogeneous model with averaged values for shear modulus and viscosity from Fig.~\ref{plot_profile_mu_eta_T1_outer_planets} (see Table~\ref{table1} for the parameters of the planets). To ease the comparison, we recall that Earth dissipates a total energy of 43-46~TW and Io between 60 and 170~TW (see main text).}
\label{table3}
\end{table*}

Table~\ref{table3} summarizes the averaged quantities: total energy dissipated (called ``Dissipation'' in the Table) and the resulting tidal heat flux at the surface of the planet for the multi-layer planet model and the homogeneous model (using the averaged shear modulus and viscosity displayed in Fig.~\ref{plot_profile_mu_eta_T1_outer_planets}).
For comparison, we calculate that the corresponding tidal dissipation in the solid Earth is 0.1~TW (which corresponds to a flux of $\sim 2\times10^{-4}$~W/m$^2$), which is similar to the dissipation in TRAPPIST-1g.  
This shows that for the Earth tidal heating is a tiny fraction of the total budget. 
Indeed, the total heat of the Earth (due to radiogenic power and secular cooling) is estimated between 44 and 46 TW \citep{1993RvGeo..31..267P, Jaupart2007}. 
The total power produced by radiogenic heat sources in the Earth's mantle and crust is estimated between 17 and 23 TW \citep{Jaupart2007}.
We also calculated the global tidal dissipation for Venus to be $\sim 3\times10^{-4}$~TW (corresponding to a flux of $\sim6\times 10^{-7}$~W/m$^2$) for the ``cold'' profile and about 10 times higher for the hot profile with a dissipation of $\sim 3\times10^{-3}$~TW.
Similarly, Io's total dissipation is estimated from both thermal emission data \citep{2000Sci...288.1198S,2004Icar..169..264V,2004Icar..169..127R} and astrometric data \citep{2009Natur.459..957L} to range between  60 and 170~TW according to observational uncertainties, with average values reported by various studies of the order of 100 TW.

We find a dissipation of $\sim 10$~TW for TRAPPIST-1e, which is 4 orders of magnitude lower than the value obtained in Table~2 of \citet{2018ApJ...857..142M}.
The dissipation we find for a homogeneous planet with averaged shear modulus and viscosity is about 3 orders of magnitude higher than for the multi-layer model. 
It is therefore closer to the estimates obtained by \citet{2018ApJ...857..142M}.
However, they have a very different approach: they are computing the maximum tidal heating rate for a homogeneous planet and a Maxwell rheology (maximized by an unphysically small Maxwell time < 1 day), whereas we are calculating the heating rate for rheological parameters consistent with present-day knowledge on high-pressure ices and rocks and realistic  internal structure models.
Figure~\ref{Comp_Trappist1e_high_iron_ice} shows that the excitation frequency of TRAPPIST-1e in our model is far from the frequency corresponding to the maximum dissipation.

For planets g and h, the total tidal power remains much smaller that the endogenic power of the Earth, indicating that it will have a negligible impact on the thermal evolution of these planets. 
For planets e and f, it becomes comparable to the present-day radiogenic power of the Earth. 
As these two planets have a rock mass estimated to about 0.7-0.8 $M_\oplus$ (Table~\ref{table1}) and are likely older than the Earth (system age estimated to $7.6\pm 2.2$ Gyr, \citealt{2017ApJ...845..110B}), the radiogenic power is likely smaller than 10 TW, making tidal heating the main heat source in these planets.
Moreover, the fact that the tidal energy is concentrated in the high-pressure ice layer may have major impacts on their internal dynamics.  
As it has been proposed for large icy moons like Titan and Ganymede \citep{2017Icar..285..252C,2018Icar..299..133K}, heat transfer in this high-pressure ice mantle may be controlled by ice melting and meltwater transport. 
The occurence of strong tidal dissipation in these icy mantles would likely promote melting, especially at the rock-ice interface due to enhanced tidal dissipation. 
In particular, in planet f, the volumetric dissipation rate at the base of the icy mantle is above $10^{-7}$~W.m$^{-3}$, which exceeds by more than one order of magnitude the volumetric heating rate by radiogenic elements in the Earth.  
This localized heat source will increase the occurrence of ice melting and the possibility of water-rock interactions at the rock-ice interface in these planets, favoring the extraction and transport of nutriments from the rock mantle to the surface. 
These may have significant implications for their habitability of such water-rich exoplanets \citep{2016Icar..277..215N,2017SSRv..212..877N}. \\

We obtain fluxes about two times lower than those estimated in \citet{2018A&A...613A..37B} for planets e and f.
They consider multi-layer planets, but do not take into account the interior layering to compute tidal dissipation. 
For each planet, they assume constant shear modulus and viscosity for the different layers, then compute volume-weighted averaged shear modulus and viscosity to finally compute the dissipation using the formulation for a homogeneous planet.
This method is different to what we perform to calculate the dissipation of a homogeneous body using the averaged shear modulus and viscosity (see Eq.~\ref{plot_profile_mu_eta_Earth_mean_fit}).
We showed in Figs.~\ref{Fit_Earth_no_constraints_multilayer_vs_homogeneous_mean_fit} and \ref{Fit_Venus_no_constraints_multilayer_vs_homogeneous} that performing that averaging and using the homogeneous body framework leads to a serious overestimation of the dissipation.
As explained in Section~\ref{dissipation_vs_frequency}, this overestimation is even greater if we compute the dissipation from the linear volume-weighted average of the shear modulus and viscosity as in \citet{2018A&A...613A..37B}. 
Indeed, due to the fact that the viscosity varies by several orders of magnitude, a linear averaging of the viscosity gives too much weight to the outer highly-viscous layer and does not allow to take into account the impact of the low viscosity layers on the dissipation.

However, we cannot strictly compare our estimates with theirs as they take into account the dependence of viscosity on temperature. 
This will be investigated in a follow-up study.

%
%
%

\section{Conclusions}

We computed the frequency dependence of various multi-layer planetary bodies: a planet of $0.5~\Mearth$, an Earth-like planet, 2 Super-Earths of 5$~\Mearth$ and 10~$\Mearth$, and TRAPPIST-1e. 
We compared it to the dissipation of homogeneous models using volume-averaged values of shear modulus and viscosity and found that doing 
so leads to a huge overestimation of the dissipation at all frequencies. 
It is therefore crucial to calculate the dissipation consistently using a formalism taking into account the mechanical properties of each internal layer and solving consistently the mechanical coupling between each internal layer when subjected to tidal forces, like, for instance, the elastic formalism proposed by \citet{1972MetComPhy...1..217S}, which can be extended to the viscoelastic case  \citep{2005Icar..177..534T}.\\

Although considering averaged values for shear modulus and viscosity does not allow to reproduce the dissipation of the multi-layer planet in a satisfactory way, the global dissipation and global tidal parameters ($k_2, Q$) can be reasonably approximated by deriving the appropriate parameters, making implementation in orbital dynamics codes simpler. 
By using computation results obtained for multi-layered planets, we derived the parameters for a equivalent homogeneous planet that best fits the multi-layered planet results. 

The fitting procedure provided a reasonable approximation for rocky planets but remains poor for ice-rich planets. 
The presence of an ice layer leads to a second dissipation peak for higher frequencies than the peak corresponding to a rocky composition.
The frequency of the second peak corresponding to the ice layer depends on the poorly constrained viscosity of the ice. 
Increasing the viscosity of the ice leads to a shift of the peak towards the shorter frequencies where it blends with the peak corresponding to a rocky composition (of higher averaged viscosity).
This double peak feature cannot be reproduced by a homogeneous planet model, so for icy planets the full multi-layer model has to be used to compute the dissipation consistently.

Using the multi-layer planet framework, we also computed the profiles of tidal heating within the outer planets of TRAPPIST-1.
We find similar values as in \citet{2018A&A...612A..86T}, which was using a constant time lag model.
We also compare our results with \citet{2018A&A...613A..37B}, which consider the dissipation rate predicted from homogeneous interior formulation to assess the dissipation in a multi-layer planets.  
As we show in the present study, such an approach using volume averaged shear modulus and viscosity  strongly overestimate the global dissipation rate. 
A limitation in our approach is that we assume for simplicity constant viscosity with depth in each internal layer and did not take into account the coupling with surface temperature-pressure conditions \citep[e.g.][]{2019A&A...631A.103B} and thermal evolution of the interior \citep{2018A&A...613A..37B}. 
As shown in the case of Venus \citep{2017JGRE..122.1338D}, pressure and temperature dependent viscosity may have a significant impact on the prediction of tidal dissipation.
Furthermore, we do not account for the presence of surface melts (magma ocean, water ocean) in our model, which is the reason why we restricted our study to the outer planets of TRAPPIST-1. 
These aspects will be treated in follow-up studies.

\begin{acknowledgements}
The authors would like to thank the anonymous referee for her/his constructive comments.
This work has been carried out within the framework of the NCCR PlanetS supported by the Swiss National Science Foundation.
S.M. and E.B. acknowledge funding by the European Research Council through ERC grant SPIRE 647383 and by the CNES PLATO grant at CEA/DAp. 
This research has made use of NASA's Astrophysics Data System.
\end{acknowledgements}

\bibliographystyle{aa}
\bibliography{Biblio}

\end{document}